\documentclass[]{bytedance_seed}
\usepackage{makecell}

\usepackage[table]{xcolor}
\usepackage{microtype}
\usepackage{hyperref}
\usepackage{url}
\usepackage{booktabs}
\usepackage{graphicx} 
\usepackage{amsmath}
\usepackage{float}
\usepackage{enumitem}
\usepackage{tabularx}
\usepackage{listings}
\usepackage{tabularx}
\usepackage{array}
\usepackage{subcaption}
\usepackage{fvextra}
\usepackage{booktabs}
\usepackage{siunitx}
\usepackage{multicol}

\usepackage[T1]{fontenc}
\usepackage{lmodern}
\usepackage{amsmath}
\usepackage{pdflscape} % 在导言区添加

\usepackage{enumitem}

\usepackage[most]{tcolorbox}
\usepackage{listings}
\usepackage{xcolor}

\newtcolorbox{casebox}[2][]{
  enhanced,
  colback=white,
  colframe=black,
  boxrule=1pt,
  arc=3mm,
  left=6pt,right=6pt,top=6pt,bottom=6pt,
  title={#2},
  attach title to upper,   % title sits inside the box
  #1
}

\newtcblisting{diffblock}{
  listing engine=listings,
  colback=green!5,
  colframe=green!40!black,
  listing only,
  left=2mm,
  right=2mm,
  top=0.5mm,
  bottom=0.5mm,
  boxrule=0.4pt,
  sharp corners,
  enhanced,
  listing options={
    basicstyle=\ttfamily\tiny,   % 这里调字号：\tiny / \scriptsize / \footnotesize
    columns=fullflexible,
    keepspaces=true
  }
}

\newcounter{case}[section]

\usepackage{lineno}

\definecolor{darkblue}{rgb}{0, 0, 0.5}
\hypersetup{colorlinks=true, citecolor=darkblue, linkcolor=darkblue, urlcolor=darkblue}

\makeatletter
\renewcommand{\sfdefault}{lmss} % or cmss

\DeclareRobustCommand{\sffamily}{\fontfamily{\sfdefault}\selectfont}
\makeatother
\title{AInsteinBench: Benchmarking Coding Agents on Scientific Repositories}

\affiliation[1]{ByteDance Seed}
\affiliation[2]{Princeton University}

\contribution{Full author list in Contributions}

\abstract{
We introduce \textsc{AInsteinBench}, a large-scale benchmark for evaluating whether large language model (LLM) agents can operate as \textit{scientific computing development agents} within real research software ecosystems. Unlike existing scientific reasoning benchmarks which focus on conceptual knowledge, or software engineering benchmarks that emphasize generic feature implementation and issue resolving, AInsteinBench evaluates models in end-to-end scientific development settings grounded in production-grade scientific repositories. The benchmark consists of tasks derived from maintainer-authored pull requests across six widely used scientific codebases, spanning quantum chemistry, quantum computing, molecular dynamics, numerical relativity, fluid dynamics, and cheminformatics. 
All benchmark tasks are carefully curated through multi-stage filtering and expert review to ensure scientific challenge, adequate test coverage, and well-calibrated difficulty.  By leveraging evaluation in executable environments, scientifically meaningful failure modes, and test-driven verification, AInsteinBench measures a model’s ability to move beyond surface-level code generation toward the core competencies required for computational scientific research. }
\date{\today}
\checkdata[Project Page]{\url{https://github.com/ByteDance-Seed/AInsteinBench}}

\begin{document}
\maketitle

\begin{figure}[H]
  \centering
  \vspace{0cm}
  \includegraphics[
    width=0.8\linewidth,
    trim=0 1.55cm 0 0,
    clip
  ]{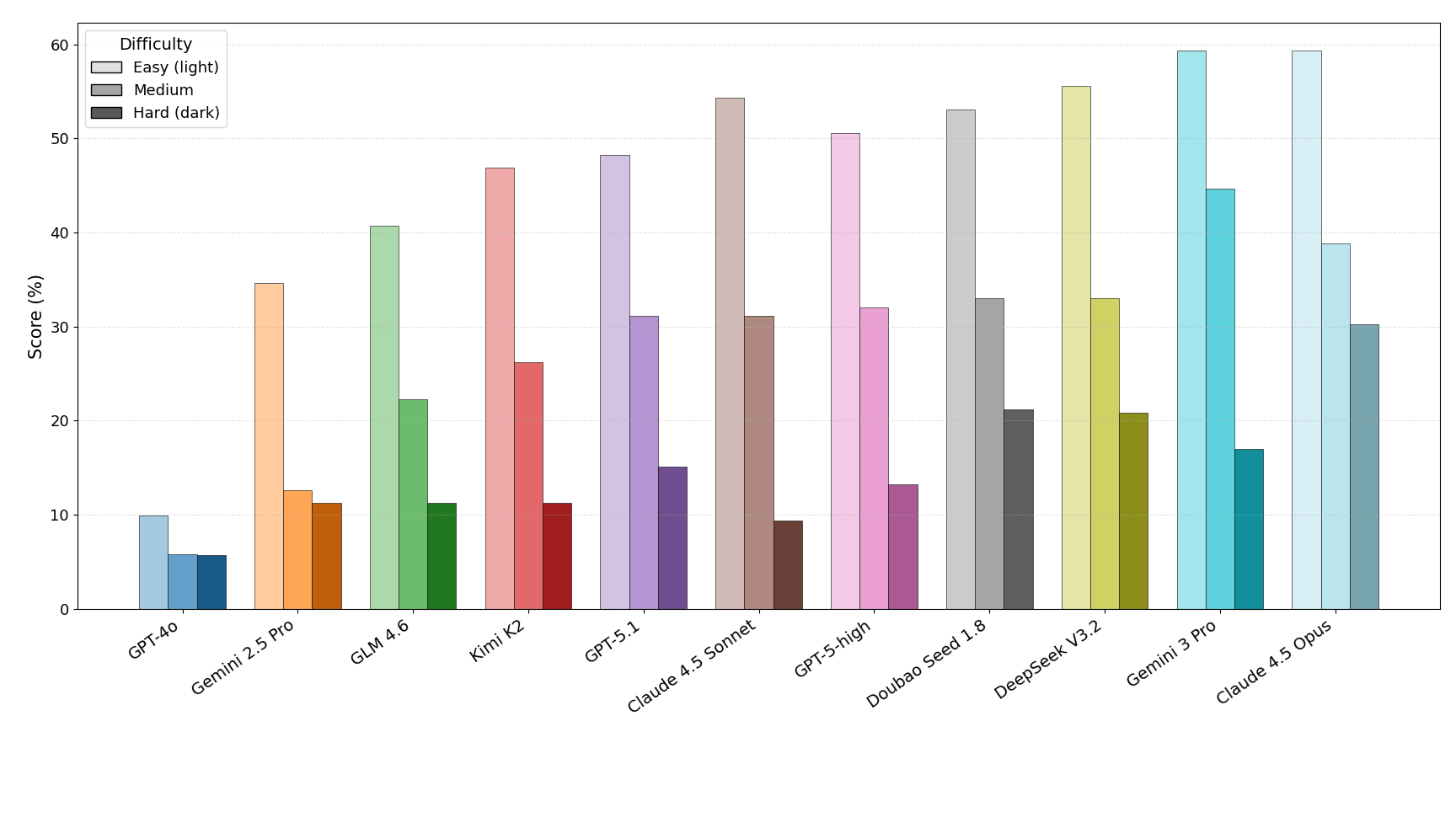}
  \caption{Overall performance score across frontier models among all difficult tiers}
  \label{fig:scicobench_cover}
\end{figure}

\section{Introduction}

\begin{figure}
    \centering
    \includegraphics[width=\linewidth]{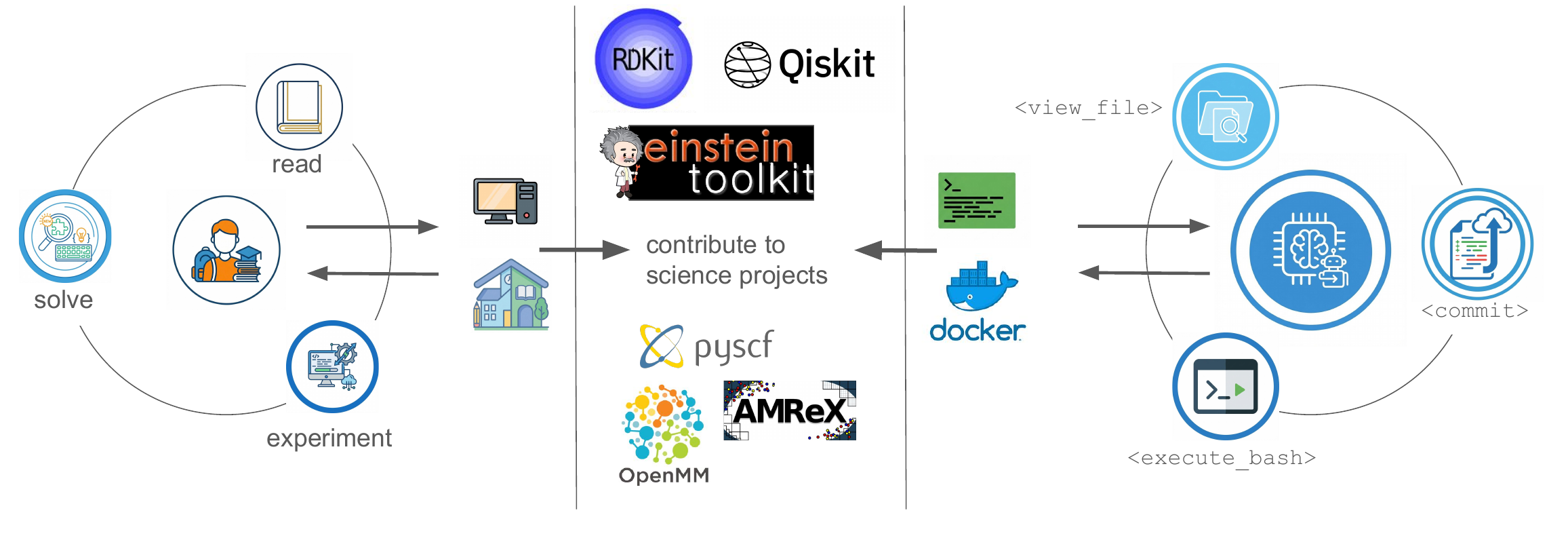}
    \caption{\textbf{Scientist vs LLM agent workflow} In computational sciences like physics, chemistry and biology, the daily work of a graduate student and postdoc usually consists of such a process: \textit{\textbf{Reading}} papers and textbooks for understanding of both science and coding; \textit{\textbf{Experimenting}} with computer simulations or actual machines or laboratories to gain feedback and iterate; \textit{\textbf{Solving}} problems or implementing new features to a project or the computational code repository it self. Throughout the process the scientist would actively interact with the computing allocations and the school/laboratory resources. On the other hand, in this paper, we setup docker container with terminal access as an environment for the LLM agent to interact with. Allowing the LLM agent to \textbf{\textit{view files}}, \textbf{\textit{execute bash}} commands, and \textbf{\textit{commit}} their changes. Through this processes, the human researches and the LLM agents contribute their ability to solve real-world problems in scientific computing projects across different disciplines including RDKit, Qiskit, EinsteinToolkit, PySCF, OpenMM, and AMReX. }
    \label{fig:scicobench}
\end{figure}

Modern scientific discovery is increasingly mediated by large-scale computational systems. 
Across disciplines such as chemistry, physics, and materials science, new scientific insights are no longer derived solely from analytical derivations or isolated simulations, but from sustained interaction with complex scientific software ecosystems. Researchers routinely formulate hypotheses, encode physical or mathematical assumptions into simulation code, and iteratively refine implementations based on empirical feedback from tests, solvers, and large-scale simulation runs. As illustrated in Figure~\ref{fig:scicobench}, such a workflow follows an iterative cycle of \textit{reading literature, experimenting,} and \textit{solving questions}, with active interactions with the environment: the academic institution and compute infrastructure. In this setting, scientific discoveries depend not only on the literal understanding of the subject, but also on the ability to implement the theory into production codebases and execute in mass production.

Recent advances in large language models (LLMs) have growing interest in their potential to assist scientific computing research \cite{Dhruv_2025_LLM,boiko2023emergentautonomousscientificresearch,yang2025largelanguagemodelsmaterials}. To function as genuine research assistants, LLM-based agents must be able to navigate long-lived scientific repositories, understand domain-specific invariants and numerical constraints, and implement or debug code in ways that preserve scientific correctness. 
Whether current models are capable of the transition from surface-level scientific reasoning toward more analytic and physically based reasoning remains an open question. As shown in Fig.~\ref{fig:scicobench}, analogous to human researchers, the LLM agents engage in a workflow of loading files into context, executing bash commands, and committing changes, with active interaction with its environment: usually a Docker container we built.

Current scientific reasoning benchmarks evaluate high-level conceptual understanding \cite{sun2024scieval,rein2023gpqa,chung2025tpbench,xu2025physense,pan2025cmtbench} or isolated coding questions \cite{tian2024scicodebench}, but they stop short of requiring LLMs to navigate, interpret, and modify real scientific repositories.  As a result, these benchmarks fundamentally underrepresent the skills needed to be a successful scientist. Across nearly all of academia, there are numerous groups whose primary workflow entails developing ideas within large scientific repositories. This is no better demonstrated than by looking through the author lists of popular scientific repositories \citep{AMReX,astropy,EinsteinToolkit,pyscf,qchem6,terrachem,openmm,qiskit,rdkit}. Existing SWE benchmarks \citep{jimenez2024swebench, zan2025multiswebench} are not much better, as they do not reflect the coding paradigms, numerical stability constraints, cross-language heterogeneity, and HPC execution environments that dominate scientific computing. Although SWE-bench includes one scientific repository, Astropy, this is by no means a representative sampling. As a result, a fundamental gap remains: no benchmark currently measures an LLM's ability to solve issue or write new features through patch synthesis inside real, large-scale scientific software ecosystems.

% To address this gap, we introduce a new benchmark, a scientific extension of Multi-SWE-bench—focused entirely on large, production scientific repositories spanning quantum chemistry (PySCF), quantum computing (Qiskit), cheminformatics (RDKit), fluid dynamics (AMReX), numerical relativity (Einstein Toolkit), and molecular dynamics simulation (OpenMM). These repositories differ significantly from standard software projects: they involve deep domain level scientific knowledge. For example, understanding the structure of density-functional theory (DFT) in quantum chemistry, the formulation of symplectic integrators in molecular dynamics, or the numerical discretizations underlying relativistic hydrodynamics and PDE solvers. Curating tasks from such environments not only increases task difficulty but also more closzely aligns benchmark design with the realities of scientific programming.

To address this gap, we introduce \textsc{AInsteinBench}, a new benchmark designed to evaluate performance in \textit{scientific computing development as practiced in real research environments}. 
Rather than extending existing software-engineering benchmarks, AInsteinBench is constructed to measure how well an agent can operate as a scientific researcher working through code: developing, diagnosing, and maintaining computational implementations that embody modern scientific theory. The benchmark is grounded in large, production-grade scientific repositories spanning quantum chemistry (PySCF), quantum computing (Qiskit), cheminformatics (RDKit), fluid dynamics (AMReX), numerical relativity (Einstein Toolkit), and molecular dynamics simulation (OpenMM). Unlike standard software projects, these codebases encode domain-specific scientific structure: from density-functional theory and self-consistent field formulations to symplectic integration, constrained optimization, and PDE discretizations.  Successfully modifying such systems requires reasoning not only about code, but about the scientific assumptions, invariants, and numerical behavior. By curating tasks from these environments, AInsteinBench moves towards whether models can translate scientific understanding into correct and stable computational practice. 
This focus aligns the benchmark with the realities of scientific programming, where progress is measured by preserving physical correctness and numerical robustness.

We construct this benchmark through a rigorous multi-phase pipeline inspired by SWE-bench but extended to meet scientific reproducibility standards. First, we identify high-impact scientific repositories with active developer communities and well-maintained test suites, ensuring both scientific relevance and executability. Second, we collect scientifically meaningful issue-linked pull requests and extract before/after code states, environment specifications, and domain-specific metadata. Third, we build fully containerized execution environments, capturing dependencies such as MPI, domain-specific compilers, and scientific development stacks. Fourth, we apply a strict filtering criterion which isolate issues that exhibit clear numerical or logical bugs with reproducible failing tests. Finally, we perform domain-expert manual verification to eliminate unsuitable instances.

% By grounding the benchmark in real scientific computation, this dataset provides several novel benefits absent from prior work:
% \begin{itemize}
% \item Scientific fidelity — tasks reflect real numerical bugs, symmetry-breaking issues, convergence failures, and algorithmic correctness problems found in production scientific codes.
% \item Cross-language complexity — scientific repositories integrate Python, C, C++, Fortran \textbf{together} expanding the scope of multi-language evaluation beyond typical software.
% \item Rich contextual structure — tasks require understanding scientific intent, mathematical formulations, numerical stability, and multi-file algorithmic interactions.
% \end{itemize}
% Together, these characteristics enable the first systematic evaluation of LLM performance on real-world scientific issue resolving, bridging the gap between conceptual scientific reasoning benchmarks and full-scale scientific software development. We further benchmark several state-of-the-art agents and frontier models, providing insights into the challenges posed by scientific repositories—including their brittle numerical behavior, tightly coupled algorithms, and long-context code semantics. Through this work, we aim to establish a foundation for future agents capable of assisting or even accelerating scientific software development at scale.

\textbf{Contributions.} Our contributions are threefold:
\begin{itemize}[leftmargin=*, itemsep=0.25em, topsep=0.3em, parsep=0pt]
    \item We formalize \textit{scientific computing as a research capability} and introduce \textsc{AInsteinBench}, a benchmark that evaluates whether LLM agents can translate scientific understanding into correct and numerically stable computational practice.
    
    \item The dataset is explicitly difficulty-calibrated and spans tasks ranging from the implementation of new computational modules requiring scientific and mathematical reasoning to the identification and correction of known repository issues, including numerical instabilities, solver non-convergence, violated invariants, and cross-language integration errors.

    \item We provide a systematic evaluation of state-of-the-art models, revealing that while current agents can resolve isolated bugs, they struggle with preserving scientific invariants, coordinating multi-file changes, and maintaining correctness in complex scientific algorithms.
\end{itemize}

% - Highlight dataset scale, diversity, and goals
% - Contributions:
%   1. New dataset
%   2. New curation pipeline
%   3. Evaluation framework and metrics
%   4. Baseline LLM performance

\section{Related Work}
% - Scientific reasoning datasets (SciEval, MMLU-Science, PubMedQA, etc.)
% - Coding + agent benchmarks (HumanEval, MBPP, SWE-Bench, DS-1000)
% - Domain-specific benchmarks (e.g., physics, bioinformatics)

\textbf{Benchmarks for mathematical and scientific reasoning.}
Evaluation of LLM reasoning in mathematics and the natural sciences has traditionally focused on ``closed-world'' problems with unambiguous targets, such as contest-style questions or short-form exam items.
In mathematics, datasets like GSM8K and MATH probe grade-school through early undergraduate problem solving, while complementary work evaluates formal reasoning via proof assistants and automated theorem proving (e.g., Lean)~\cite{cobbe2021training,hendrycks2021measuring,polu2020generative}.
As frontier models improve, these settings increasingly face saturation and data-contamination concerns, motivating benchmarks that rely on newly written, held-out problems, scalable scoring procedures, and explicit difficulty calibration~\cite{glazer2024frontiermath}.
Recent systems have reached olympiad-level performance in geometry~\cite{trinh2024solving} and begun to support research-style advances via program search~\cite{romera2024mathematical}, highlighting the limits of standard competition corpora.
Reflecting this shift, FrontierMath introduces expert-authored, unpublished problems with automated verification and structured difficulty ratings~\cite{glazer2024frontiermath}, while in the sciences, GPQA raises conceptual difficulty but remains question-answering–oriented~\cite{rein2023gpqa}.
FrontierScience pushes further toward research-oriented assessment by separating an Olympiad track from a rubric-graded Research track of domain-authored tasks~\cite{openai2025frontierscience}.

\textbf{Foundation Models for Science.}
A growing line of work aims to endow language models with stronger scientific priors through domain-targeted pre-training and instruction tuning, without introducing explicit agentic control loops.
Early efforts emphasized scientific and biomedical language modeling via in-domain pre-training (e.g., SciBERT, BioBERT, PubMedBERT), improving scientific NLP as a substrate for downstream reasoning and knowledge use~\cite{beltagy2019scibert,lee2020biobert,gu2020pubmedbert}.
More recent work extends this paradigm to \textit{generative} scientific foundation models, including domain-specialized biomedical generators such as BioGPT and broader science-focused LLMs such as Galactica, which are trained on large scientific corpora and evaluated on scientific QA, knowledge-intensive tasks, and technical text generation~\cite{luo2022biogpt,taylor2022galactica}.
In parallel, instruction- and adaptation-based approaches build explicit scientific instruction data or specialized scientific LLMs: \cite{zhang2024sciglm} propose \textsc{SciInstruct} and report gains in college-level scientific reasoning, \cite{zhao2024chemdfm} introduce \textsc{ChemDFM} for chemistry-centric QA and molecular reasoning, and \cite{sun2024scidfm} develop \textsc{SciDFM}, a mixture-of-experts scientific LLM trained at large scale.
% Overall, these efforts improve models' \textit{in-domain priors} and static benchmark performance, but they typically do not test whether models can maintain scientific and numerical correctness while operating inside real scientific software ecosystems.

\textbf{LLM Agent for Science.}
A complementary line of research studies LLM-based agents that couple language reasoning with tools, simulators, or multi-step interaction loops to perform scientific tasks beyond standalone question answering.
Representative systems integrate domain tools for chemistry and materials science~\cite{bran2024chemcrow,boiko2023emergentautonomousscientificresearch}, or evaluate agents as biomedical hypothesis generators that iteratively propose, critique, and refine hypotheses from background evidence~\cite{qi2024biomedhypo}.
Recent benchmarks further connect agents to executable scientific simulators, enabling iterative \textit{in silico} experimentation and feedback-driven reasoning~\cite{duan2025scigym}.
While these works emphasize planning--acting--observing loops and experimental orchestration, they largely focus on hypothesis generation or simulation control, rather than repository-scale scientific software development and code-level reasoning.

\textbf{Agentic Coding Benchmarks.} Recent efforts have introduced increasingly realistic benchmarks for evaluating
LLMs on software engineering tasks. SWE-Bench
formulates issue resolution as patch generation over real Python repositories \cite{jimenez2023swebench}.
Its extension, SWE-Bench Verified, incorporates
human validation to improve alignment between issues and reference patches,
although the setting remains monolingual and primarily oriented toward small,
localized bug fixes \cite{jimenez2023swebenchverified}. Building on this line of work, SWE-Gym
converts SWE-Bench tasks into an interactive environment, enabling multi-step
agentic editing rather than single-shot patch generation, while inheriting the
same task distribution and Python-only scope \cite{kulal2024swegym}. More recently, Multi-SWE-Bench expands the evaluation to a multilingual setting,
introducing issue-resolution tasks across seven widely used programming
languages \cite{zan2025multiswebench}.

\textbf{The Need for a Benchmark for Scientific Computing Agents.}
Prior evaluation of LLMs for science largely falls into two regimes: closed-world scientific reasoning and QA benchmarks (e.g., GPQA and related expert-level test sets)~\cite{rein2023gpqa,phan2025humanitys,sun2024scieval}, and agentic systems that close the loop via tools or simulators (e.g., SciGym and hypothesis-generation frameworks)~\cite{duan2025scigym,qi2024biomedhypo}.
In parallel, software-engineering benchmarks such as SWE-Bench, SWE-Bench Verified, SWE-Gym, and Multi-SWE-Bench evaluate repository-based issue resolution~\cite{jimenez2024swebench,jimenez2023swebenchverified,kulal2024swegym,zan2025multiswebench}, but they overwhelmingly reflect general-purpose software and underrepresent the required knowledge and skills in scientific computing.
As a result, existing evaluations do not directly measure whether agents can perform the core computational work of modern research: navigating, modifying, and validating production scientific repositories where correctness is scientific and numerical, not merely syntactic or local.
\textsc{AInsteinBench} is introduced to fill this missing intersection.

\section{Dataset}

In this section. we explain the process that we produce, select and verify questions in the benchmark. We first list the scientific computing projects that we select and why we select them. Then two categories of tasks, feature implementation and issue resolving, are defined on the projects. After that, we explain the details of how we produce questions in Data Curation. Finally, the questions are scrutinized and verified by human experts we recruited.

\begin{table}[t]
\centering
\small
\setlength{\tabcolsep}{5pt}
\renewcommand{\arraystretch}{1.1}

\begin{tabularx}{\linewidth}{@{} l p{3.3cm} X @{}}
\toprule
\textbf{Repo} & \textbf{Domain/Role} & \textbf{Features} \\
\midrule

PySCF &  Theoretical Quantum Chemistry
      & High-accuracy electronic-structure solvers; modular Python interfaces. \\

Qiskit & Quantum Computing theory and infrastructure
       & Quantum circuit construction and simulation; hardware-backend execution interfaces. \\

Einstein Toolkit & Computational Astrophysics
                 & Large-scale relativistic simulations; data-calibrated pipelines connecting theory and observation. \\

AMReX & Computational Physics and Numerical Infrastructure
      & High-performance AMR grid framework; reusable solvers and mesh utilities. \\

OpenMM & Computational Molecular Dynamics
       & Force-field molecular dynamics simulations; optimized kernels for biomolecular systems. \\

RDKit & Cheminformatics
      & Molecular property prediction; cheminformatics algorithms and fingerprint utilities. \\

\bottomrule
\end{tabularx}
\caption{Overview of selected scientific repositories and their primary features.}
\label{tab:repo_overview}
\end{table}

% \subsection{Repository and Pull Request Selection}
\subsection{Project Selection}

Our benchmark focuses on scientific codebases that are widely used and representative of real research workflows. We select repositories using four criteria:

\begin{itemize}[
    leftmargin=*,
    labelsep=0.6em,
    itemsep=0.25em,
    parsep=0pt,
    topsep=0.25em,
    % partopsep=0pt
]
\item \textbf{Acceptance.} We prioritize repositories that have become de facto standards within their respective research domains. Such projects typically accumulate substantial user bases, community discussions, and long-term citations, reflecting both practical relevance and sustained scientific trust.
\item \textbf{Breadth.} We include projects that differ in disciplinary focus (e.g., quantum chemistry, numerical relativity, molecular dynamics, and fluid simulation) and in software stacks, spanning Python front ends and performance-critical C/C++/Fortran kernels.
\item \textbf{Maintenance.} We select repositories with active issue discussions, code reviews, and regular releases, and with well-structured test suites that exercise core functionality and enable reliable evaluation.
\item \textbf{Unit tests.} Each repository provides a self-contained, reproducible build-and-test workflow that can be executed at fixed commits, allowing controlled evaluation without relying on external data or environment-specific configuration.
\end{itemize}

% These criteria ensure that the selected codebases reflect real-world scientific practice while exposing models to varied domains, programming paradigms, and testing environments.

Based on these principles, we select a set of widely used scientific software repositories that satisfy all four criteria and collectively represent the diversity of modern computational research practice.
These repositories are summarized in Table~\ref{tab:repo_overview}.

\subsection{Task Description}

Each question in the benchmark represents a real scientific computing project's development task. For every task, we provide the full problem context: the issue/PR description (for repo-crawled questions), the ``before'' code state, the test patch used to expose the error, and the ``after'' code state representing the ground-truth fix. The agent interacts with this environment through file edits and test executions.

\textbf{Feature-implementation tasks.} 
FEA-Bench has addressed the question of whether LLMs are capable of incremental software development across large and heterogeneous codebases \cite{sun2025feabench}. In contrast, our benchmark focuses on a more domain-specific capability: assessing whether models can correctly implement concrete computational modules within established scientific computing repositories.
% @shuo.xin needs addition and write a \begin{casebox} for EinsteinToolkit, similar to the Qiskit case.
For instance, in our production of synthesized \verb|EinsteinToolkit| questions, we ablate a feature (a ``Thorn'' of \verb|Cactus|) by removing part of the source codes and ask LLM to implement the ablated feature by completing the missing code. In addition, a few repo-crawled questions also involve feature implementation.

\textbf{Bug-fixing tasks.}
Tasks that involve solving a science-related bug, such as physical conservation laws, group representations, calculus, and so on. The relevant bugs typically propagate across multiple files and require understanding both the science context and the software architecture. Most of the repo-crawled questions fall into this kind of task.

\begin{figure}[H]
    \centering
    \includegraphics[width=1.0\linewidth]{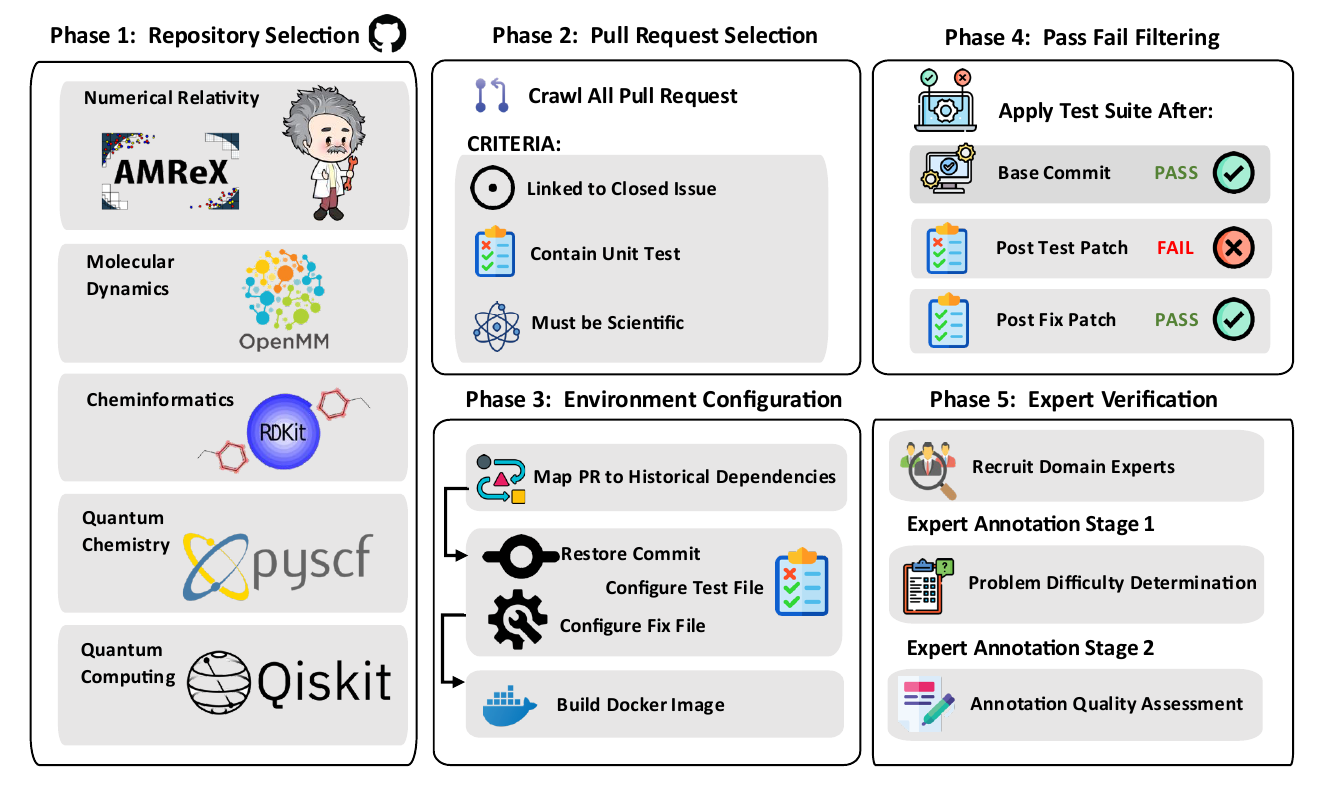}
    \caption{Overall data curation and benchmark construction pipeline.}
    \label{fig:pipeline}
\end{figure}

\subsection{Data Curation}

The questions that we include in our benchmark generally are produced by two kinds of pipelines: \textit{Synthesized Questions}, where we ablate some functionality of a code base and ask LLM to implement, and \textit{Repo-crawled Questions}, where one question comes from a pull request and related issues on GitHub.

\textbf{Repo-Crawled Questions}

\begin{itemize}[
    leftmargin=*,
    labelsep=0.6em,
    itemsep=0.25em,
    parsep=0pt,
    topsep=0.25em,
    % partopsep=0pt
]
    \item \textbf{Pull Request Crawling.}  We begin by cloning each repository and collecting all pull requests via the GitHub API, extracting metadata such as the issue description, base commit, fix patch, and associated test updates. This yields the ``before'' and ``after'' code states for potential benchmark instances.  
To ensure high-quality tasks, we retain only PRs that satisfy three criteria:  
(1) they address a clearly defined scientific or numerical bug rather than documentation, refactoring, or large feature additions;  
(2) they include test updates that expose the failure and validate the fix; and  
(3) they are merged into the main branch, indicating correctness and compatibility with the repository.  
This filtering stage produces a curated pool of scientifically meaningful, test-verifiable code changes from which benchmark tasks are constructed.\\
\item \textbf{Environment Configuration.} Software repositories evolve over time (dependencies change, APIs shift, and default behaviors vary across releases), making historical reproducibility nontrivial. To ensure robust and consistent evaluation, we reconstruct an environment for each pull request that mirrors the dependency stack at or near the time it was created.
Each instance is executed inside a minimal, isolated Docker container that installs the required libraries, configures build paths, and enables deterministic test execution. This design guarantees cross-machine reproducibility and prevents interference between repositories or dependency versions.\\
\item \textbf{FP Filtering.} Following the data-construction pipelines of SWE-Bench and Multi-SWE-Bench~\cite{jimenez2024swebench,zan2025multiswebench}, we collect PRs from actively maintained scientific repositories, retain those that resolve documented issues and modify tests, and verify that each instance produces a clear fail-to-pass transition when executed in the reconstructed environment.

\item \textbf{New Components Extraction and Feature Implementation.} Furthermore, to enhance the quality of the benchmark dataset, specifically for problems involving new feature requests, we implemented a supplementation pipeline that augments issue descriptions with precise API documentation. In many cases, when a PR introduces new functionality (features), the original issue description, or the PR description itself, lacks sufficient detail about the new API's signature and usage. This scarcity of information renders it near impossible for AI agents implement new features, since this requires guessing the new function signatures. Rather than throwing these valuable problems out, we can `supplement' the issue descriptions.

\end{itemize}

Our supplementation process focuses on ``adding docstrings'' in the form of typed API signatures to the problem description. This ensures that the agent has immediate access to the definition of new tools available in the codebase. The pipeline consists of the following steps:(1) We analyze the git diff of the PR's ``fix patch'' to identify all newly introduced function and class definitions and extract the raw signatures, including parameter names and available type annotations.(2) We filter the extracted APIs. By parsing the ``test patch'' associated with the PR, we identify which of the new symbols are actually utilized in the tests. Only these relevant APIs are selected for supplementation, ensuring the added context is directly applicable to the verification logic.(3) Fully typed API signatures are formatted into a standardized ``Minimal API'' block. This block includes the function/class location, signature, and a structured list of inputs and outputs. This documentation is appended to the original issue description in the dataset, effectively ``adding docstrings'' that describe the new feature's interface to the problem statement.

\textbf{Synthesized Questions}

\begin{itemize}[
    leftmargin=*,
    labelsep=0.6em,
    itemsep=0.25em,
    parsep=0pt,
    topsep=0.25em,
    % partopsep=0pt
]
    \item \textbf{Question Production.} We begin by identifying computational modules or features in key scientific software repositories, such as ``Thorns'' in the \verb|EinsteinToolkit|  framework. Similar workflows can also be produced analogously for other repositories. We selectively ablate a feature, typically by removing or masking parts of source code within the module (e.g., files in \texttt{src/}), and construct a task prompt based on the accompanying documentation, interface specifications, remaining code context, and problem description. Each instance is designed to simulate a realistic developer scenario in which an agent must reconstruct or implement a missing component from the intended functionality, like filling the grid with a certain spacetime metric, implementing a source term for right-hand-side of differential equations, doing spherical decomposition for finding apparent horizons, etc. An example is given in Appendix \ref{app:ET_example}.\\
    \item \textbf{Test Verification.} Since a desired functionality usually has different ways of implementation, the original code is only a reference answer but not a good verification. Instead, each feature task defined is paired with one or more test simulations defined by parameter files and output checks (e.g., reference data in \texttt{test/} directories for \verb|EinsteinToolkit|  thorns) that validate correct behavior of the implementation. An agent's solution is accepted if, when inserted in the appropriate code location, it passes all tests in the reconstructed environment, thereby ensuring that the feature behaves as intended and satisfies the scientific correctness criteria.
\end{itemize}

\subsection{Human Verification}\label{sec:Human-verf}

To ensure scientific fidelity and evaluation reliability, we perform domain-expert human verification as the final curation stage. This step serves two purposes: (i) confirming that each instance encodes a clear and meaningful scientific background, and (ii) auditing and, auditing and, when necessary, revising the associated tests to ensure adequate coverage.

\textbf{Issue Verification.} 
Each candidate instance is reviewed by domain experts to determine whether the underlying change reflects a scientifically meaningful issue. Issues that are purely engineering-driven and lack scientific relevance are excluded. Crucially, experts also assess \textit{clarity}: the issue/PR description must provide enough information for an agent to infer failure mode  and the expected behavior using only in-repo context.
Instances with ambiguous intent, missing reproduction conditions, or underspecified expected outcomes are filtered out; when the intent is sound but the description is incomplete (most commonly for feature requests), we supplement the prompt with minimal, test-relevant API documentation. 
Finally, experts assign an empirical difficulty tier (e.g., easy/medium/hard) reflecting the scientific depth and engineering complexity required to solve the instance (Please see detail in Appendix), which we later use for stratified analysis. Please see the difficulty grading rubrics in Appendix-~\ref{sec:grading-rubrics}

\textbf{Test Verification.} A benchmark is only as reliable as its tests. For each retained instance, we audit the test patch to ensure that it faithfully captures the intended scientific requirement and is robust to common shortcut solutions.
We explicitly check two failure modes and revise tests accordingly:
\textit{(i) under-coverage}: tests are too weak and admit spurious fixes. In these cases, we strengthen tests by adding regression cases.
\textit{(ii) over-coverage}: tests over-constrain the solution by encoding a particular implementation strategy, thereby rejecting alternative but scientifically valid fixes. In these cases, we remove tests that enforce irrelevant or overly prescriptive implementation details.

\textbf{False Positives.}
A \textit{false positive} occurs when a model-generated patch passes the benchmark tests but does not resolve the underlying scientific issue.
This typically arises from under-coverage, where tests validate superficial behavior (e.g., an exception path or a loose numerical bound) rather than the scientific constraint itself.
We mitigate this by strengthening tests and by adversarially inspecting candidate patches for common shortcuts; Case Study~\ref{case:chiral-fp} provides an illustrative example.

\textbf{False Negatives.}
A \textit{false negative} occurs when a patch resolves the scientific issue but fails the benchmark due to over-coverage.
We mitigate this by removing prescriptive assertions and validating scientific properties up to appropriate numerical tolerance; Case Study~\ref{case:diis-fn} illustrates how overly strict iteration caps can incorrectly reject valid fixes.

\subsection{Dataset Statistics}
\begin{figure}[htb]
    \centering
    \includegraphics[width=0.95\linewidth]{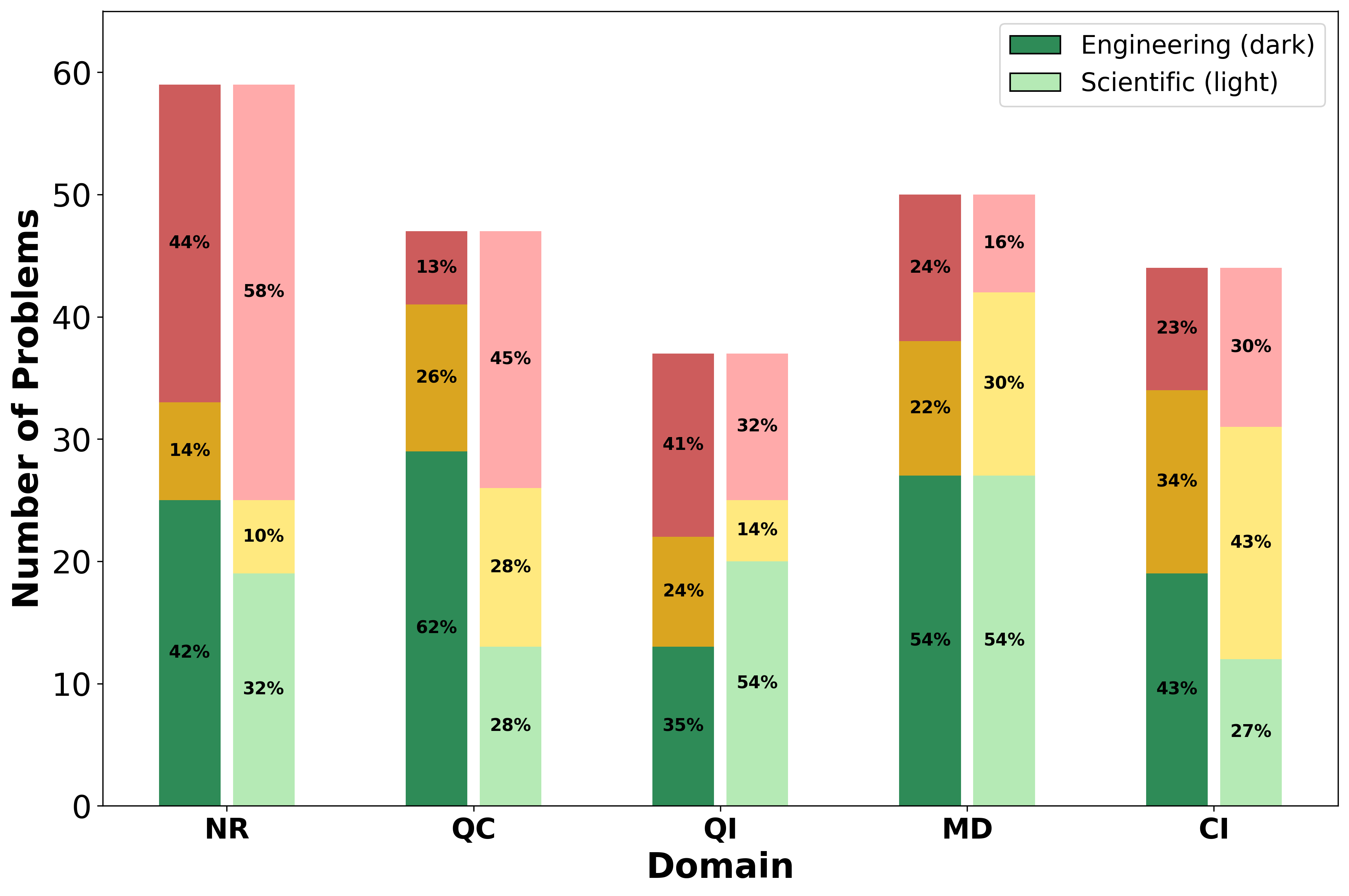}
    \caption{Scientific and engineering difficulty by domain(green: easy, yellow: medium, red: hard).}
    \label{fig:difficulty_stacked}
\end{figure}

{
In total, \textsc{AInsteinBench} contains 244 problems drawn from real pull requests and feature implementations across six repositories and multiple scientific domains.
Each instance corresponds to a maintainer-validated change gated by repository tests, yielding an evaluation suite. The difficulty tiers across the five domains are shown in Fig. \ref{fig:difficulty_stacked}. }

\section{Evaluation}
\subsection{Evaluation Methods}

% Our evaluation protocol combines a fixed set of closed-source and open-source LLMs with a custom agentic framework that mediates all interactions with the benchmark environments. 

\textbf{Coding Agent.}  We deploy the \textit{CodeAct Agent}, a modular framework that integrates LLMs with a containerized execution environment and programmatic tools~\cite{codeact}. For each benchmark instance, the agent provisions an isolated Docker container with the appropriate repository snapshot and dependencies, providing a restricted command-line interface. This setup enables the agent to inspect files, modify code, install packages, and run tests in a reproducible, sandboxed environment. The agent follows a plan–act loop: first, it gathers information (e.g., searching the codebase, inspecting test failures), then proposes concrete edits. After each edit, the agent invokes custom reproduction scripts, analyzes failures, and iteratively refines its patch, ensuring it produces changes that robustly pass the repository’s tests.

\textbf{Models.}
We evaluate a diverse set of state-of-the-art large language models, spanning both proprietary and open-source models. Specifically, we include GPT-4o, GPT-5.1, GPT-5-high, Claude 4.5 Sonnet, Claude 4.5 Opus, Grok-4, Gemini 3 Pro, Gemini 2.5 Pro, and Doubao Seed 1.8. We additionally report results for open-source models Kimi K2, GLM 4.6, and DeepSeek V3.

\textbf{Evaluation Metric.} Given the structure of this dataset, the natural evaluation metric is the proportion of tasks which are successfully addressed by the agent's patch. Beyond the basic problem solve rate, we evaluate the models ability on the following metrics, (1) \textit{scientific domain}, measuring variation across physics-, chemistry-, and engineering-focused repositories;  
(2) \textit{issue localization}, assessing whether models edit the correct files implicated by a bug;  
(3) \textit{problem difficulty}, evaluating robustness across empirically stratified tiers; and  
(4) \textit{efficiency}, quantifying efficiency finding, fixing/implementing, and testing the prescribed issue/feature. . Together, these axes highlight both structural challenges in scientific codebases and systematic strengths and weaknesses of current LLM-based coding agents.

% We evaluate performance on \textsc{AInsteinBench} along five dimensions:  
% (1) \textit{scientific domain}, measuring variation across the scientific domains in our repository;  
% (2) \textit{issue localization}, assessing whether models edit the correct files implicated by a bug;  
% (3) \textit{problem difficulty}, evaluating robustness across empirically stratified tiers; and  
% (4) \textit{efficiency}, quantifying efficiency finding, fixing/implementing, and testing the perscribed issue/feature.  
% Together, these dimensions highlight both structural challenges in scientific codebases and systematic strengths and weaknesses of current LLM-based debugging agents.

\subsection{Model Performance}

% Preamble 里加上（如果你还没加）：
% \usepackage{booktabs}
% \usepackage{siunitx}
% \sisetup{
%   detect-weight=true,
%   detect-family=true,
%   table-number-alignment=center
% }

\begin{table}[t]
\centering
\label{tab:model_results}
\footnotesize
\setlength{\tabcolsep}{4pt}
\renewcommand{\arraystretch}{1.15}
\begin{tabular}{l *{9}{S[table-format=2.1]}}
\toprule
Model
& {\%Resolved$\uparrow$}
& \multicolumn{3}{c}{Difficulty}
& \multicolumn{5}{c}{Project} \\
\cmidrule(lr){3-5}\cmidrule(lr){6-10}
&  & {Hard} & {Medium} & {Easy} & {NR} & {QI} & {MD} & {QC} & {CI} \\
\midrule

\textbf{Claude 4.5 Opus}
& \bfseries 44.0
& \bfseries 30.2 & 38.8 & \bfseries 59.3
& 44.1 & \bfseries 67.6 & \bfseries 53.7 & 38.3 & \bfseries 19.6 \\

Gemini 3 Pro
& 42.8
& 17.0 & \bfseries 44.7 & \bfseries 59.3
& \bfseries 45.8 & 62.2 & 51.9 & 36.2 & \bfseries 19.6 \\

DeepSeek V3.2
& 37.0
& 20.8 & 33.0 & 55.6
& 33.9 & 63.1 & 46.2 & 34.0 & 10.8 \\

Doubao Seed 1.8
& 36.7
& 21.2 & 33.0 & 53.1
& 34.5 & 54.1 & 40.7 & \bfseries 40.4 & 17.4 \\

GPT-5-high
& 35.4
& 13.2 & 32.0 & 50.6
& 35.6 & 54.1 & 40.7 & 34.0 & 8.7 \\

Claude 4.5 Sonnet
& 33.7
& 9.4 & 31.1 & 54.3
& 37.3 & 48.6 & 35.2 & 34.0 & 15.2 \\

GPT-5.1
& 32.9
& 15.1 & 31.1 & 48.2
& 35.6 & 51.4 & 35.2 & 34.0 & 10.9 \\

Kimi K2
& 29.1
& 11.3 & 26.2 & 46.9
& 35.5 & 51.3 & 31.5 & 25.5 & 4.3 \\

GLM 4.6
& 25.9
& 11.3 & 22.3 & 40.7
& 37.3 & 37.8 & 25.9 & 27.7 & 0.0 \\

Gemini 2.5 Pro
& 19.3
& 11.3 & 12.6 & 34.6
& 22.0 & 35.1 & 20.4 & 14.9 & 6.5 \\

GPT-4o
& 7.0
& 5.7 & 5.8 & 9.9
& 13.6 & 16.2 & 3.7 & 2.1 & 0.0 \\

\bottomrule
\end{tabular}
\caption{Models evaluated on the benchmark for each domain with resolved rates. Difficulty is reported as pass rate (\%) over easy, medium, and hard instances. The five scientific domains and corresponding codebases are: Numerical Relativity (NR): EinsteinToolkit and AMReX; Quantum Information (QI): Qiskit; Molecular Dynamics (MD): OpenMM; Quantum Chemistry (QC): PySCF; Cheminformatics (CI): RDKit.}
\label{table:model-repo-performance}
\end{table}

\textbf{Performance Across Scientific Domains.} Table \ref{table:model-repo-performance} summarizes resolution rates across repositories and scientific fields. We observe substantial variation across domains. Cheminformatics-oriented repositories such as \textsc{RDKit} and HPC-focused repositories such as \textsc{AMReX} exhibit lower solve rates. In the case of \textsc{AMReX}, this is due to significant engineering difficulty of the repositories pulled issues. For RDKit, the low resolve rate stems from  the demand for genuine spatial reasoning, a capability where current models still lag behind \citep{ijcai2025p1200}. In particular, a large fraction of RDKit tasks hinge on correctly interpreting subtle aspects of a molecule’s 3D structure. In contrast, repositories like \textsc{Einstein Toolkit} benefit from rich in-repo tooling and testing harnesses, and hinge much more on correct mathematical implementations of scientific equation. This reduces agent burden, yielding higher success rates.

\textbf{Performance Across Problem Difficulty.}

\begin{figure}[htb]
    \centering
    \includegraphics[width=0.8\linewidth]{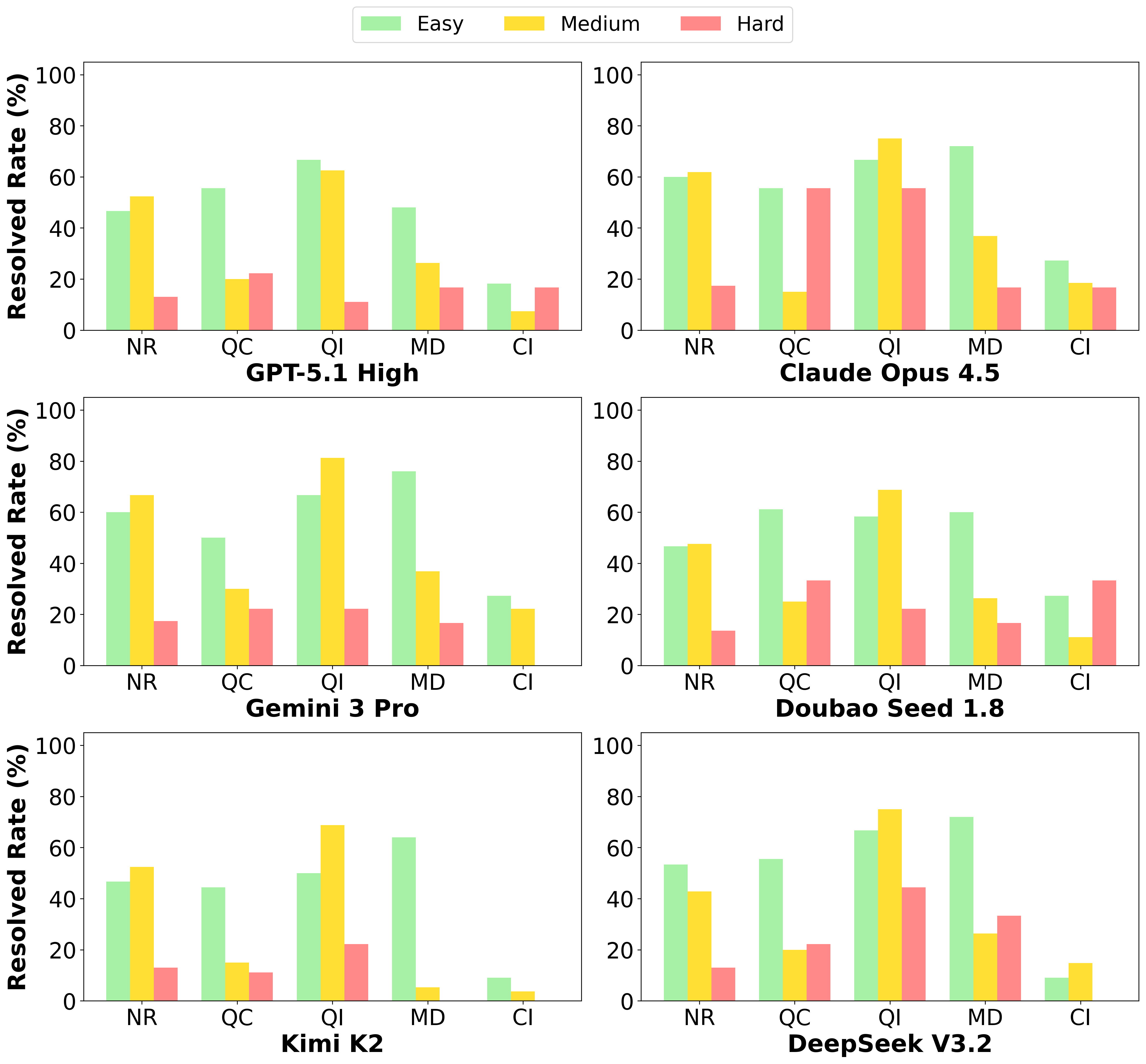}
    \caption{Problem resolution rate across difficulty tiers and scientific domains
(green: easy, yellow: medium, red: hard). Overall difficulty is defined as the mean of engineering and scientific difficulty}
    \label{fig:solve_rates_by_domain_difficulty_avg}
\end{figure}

We assess the detailed problem-resolved-rate distribution across difficulty. We bucket the difficulties as easy $[1,\,2.\overline{33}]$, medium $(2.\overline{33},\,3.\overline{66}]$, and hard $(3.\overline{66},\,5]$. Overall, the performance decreases monotonically with difficulty 
% (Figure~\ref{fig:difficulty-performance})
Fig.~\ref{fig:solve_rates_by_domain_difficulty_avg}. 
% Hardest-tier examples are dominated by XXX. 
The difficulty-stratified results incorporate the human review from recruited domain experts, whose feedback is based not only on the question statement, but also samples of trajectories of LLM agents solving the question in the environment. For each question, the difficulty is not only marked by a score, but also a feedback in text so that different experts can cross validate.

\subsection{Statistics and Analysis}

\begin{figure}[htb]
    \centering
    \begin{subfigure}[b]{0.45\textwidth}
        \centering
        \includegraphics[width=\linewidth]{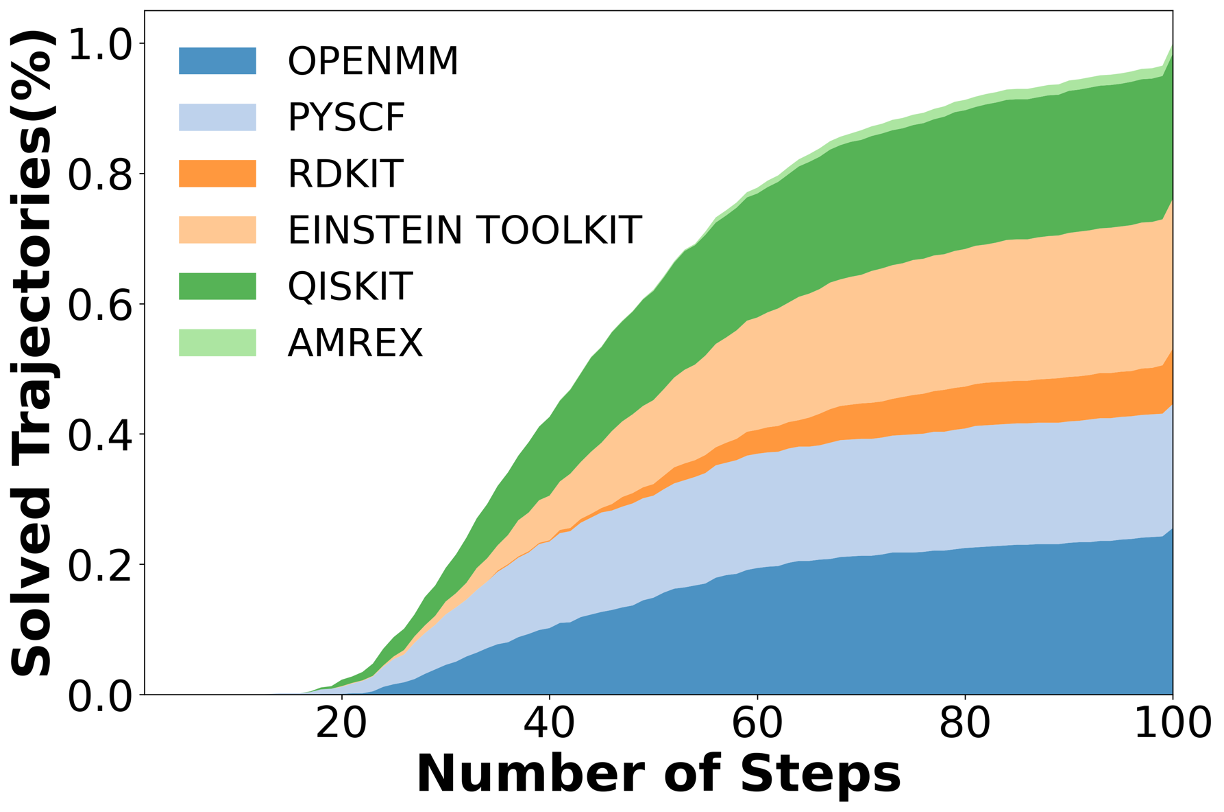}
        \label{fig:success_stack}
    \end{subfigure}
    \begin{subfigure}[b]{0.45\textwidth}
        \centering
        \includegraphics[width=\linewidth]{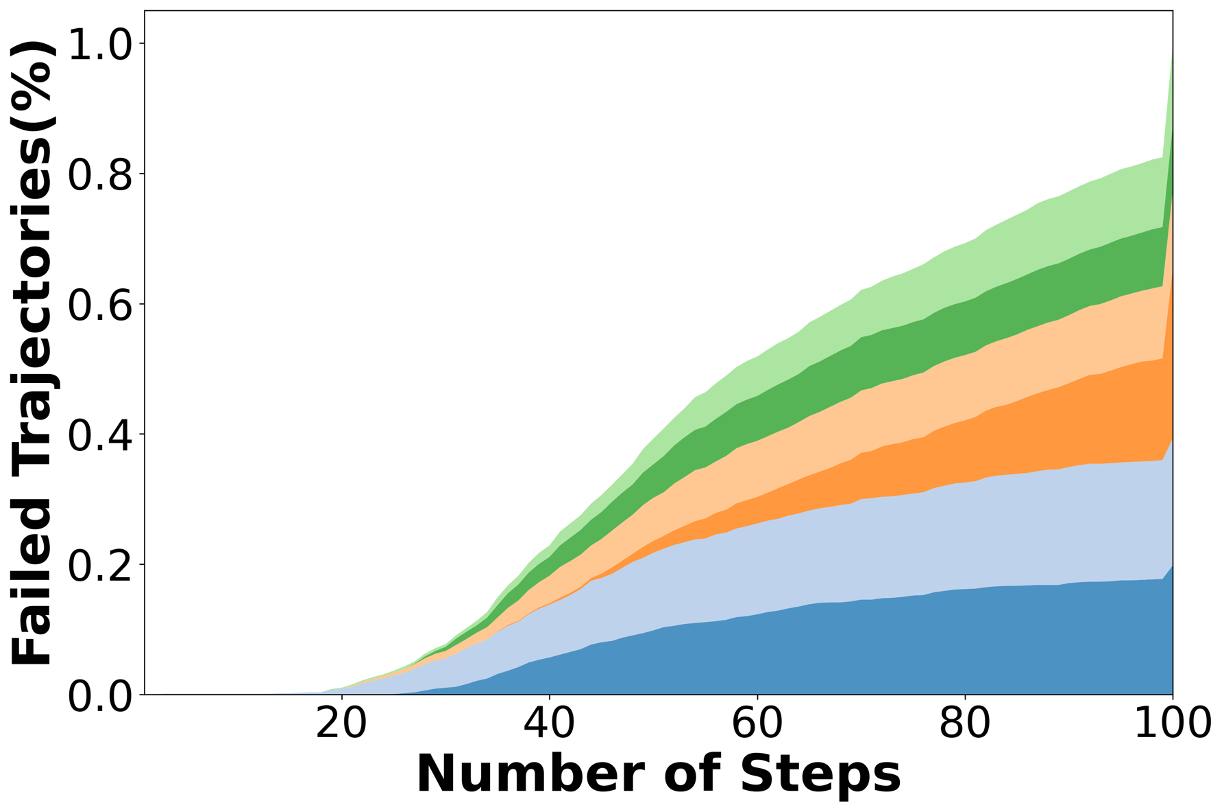}
        \label{fig:failure_stack}
    \end{subfigure}
    \caption{Cumulative solved (left) and failed (right) trajectories versus number of iterations across scientific repositories. Solved trajectories saturate early, while failures accumulate steadily with additional steps.}
    \label{fig:iteration-stack}
\end{figure}

\textbf{Efficiency and Search Cost.} During the agent's work, each additional round of conversation not only incurs computational cost but also exploits the model's context window and increases the risk of compounding errors\cite{liu2023lostmiddlelanguagemodels}. In order to quantify the efficiency, we measure the number of iterations each method requires to reach a terminal outcome (success or failure) (Figure~\ref{fig:iteration-stack}).

The trajectory-level breakdown reveals that virtually all problems take at least 20 iterations, a startup period in which the models are exploring the structure of the repository, localizing the problem, and attempting to recreate it. Most solutions accumulate quickly in the first 20-60 steps, after which point additional steps show only moderate improvement. Beyond this point, additional steps primarily contribute to failed trajectories rather than converting failures into successes\cite{yang2024sweagentagentcomputerinterfacesenable}. This pattern suggests that success depends on satisfying a small number of tightly coupled global conditions, such as identifying the correct files and implementing a fix that is consistent across all affected components. In contrast, trajectories that fail to meet these conditions early are rarely able to correct their mistakes transition into successful solutions, leading to a steady accumulation of unsuccessful attempts as step budgets increase.

% How did the different models perform across domain, language, difficulty. 
% How did different agents perform
% Convergence success across trial number(Shuo's results)
% How do models perform at locating issue

\textbf{Issue Localization.} Fixing scientific software often requires identifying \textit{where} in a large, multi-language codebase the underlying defect resides. We evaluate localization independently of correctness by determining whether a candidate patch interacts with the files required files. Here `interacts' is defined as editing, viewing the contents of, or deleting the file in question. Due to the complexity of these patches, and the existence of redundant files (such as Readme.md, certain header files, etc.) we require a pipeline for determining what 'required' changes are.

\begin{figure}[htb]
    \centering
    \includegraphics[width=1.0\linewidth]{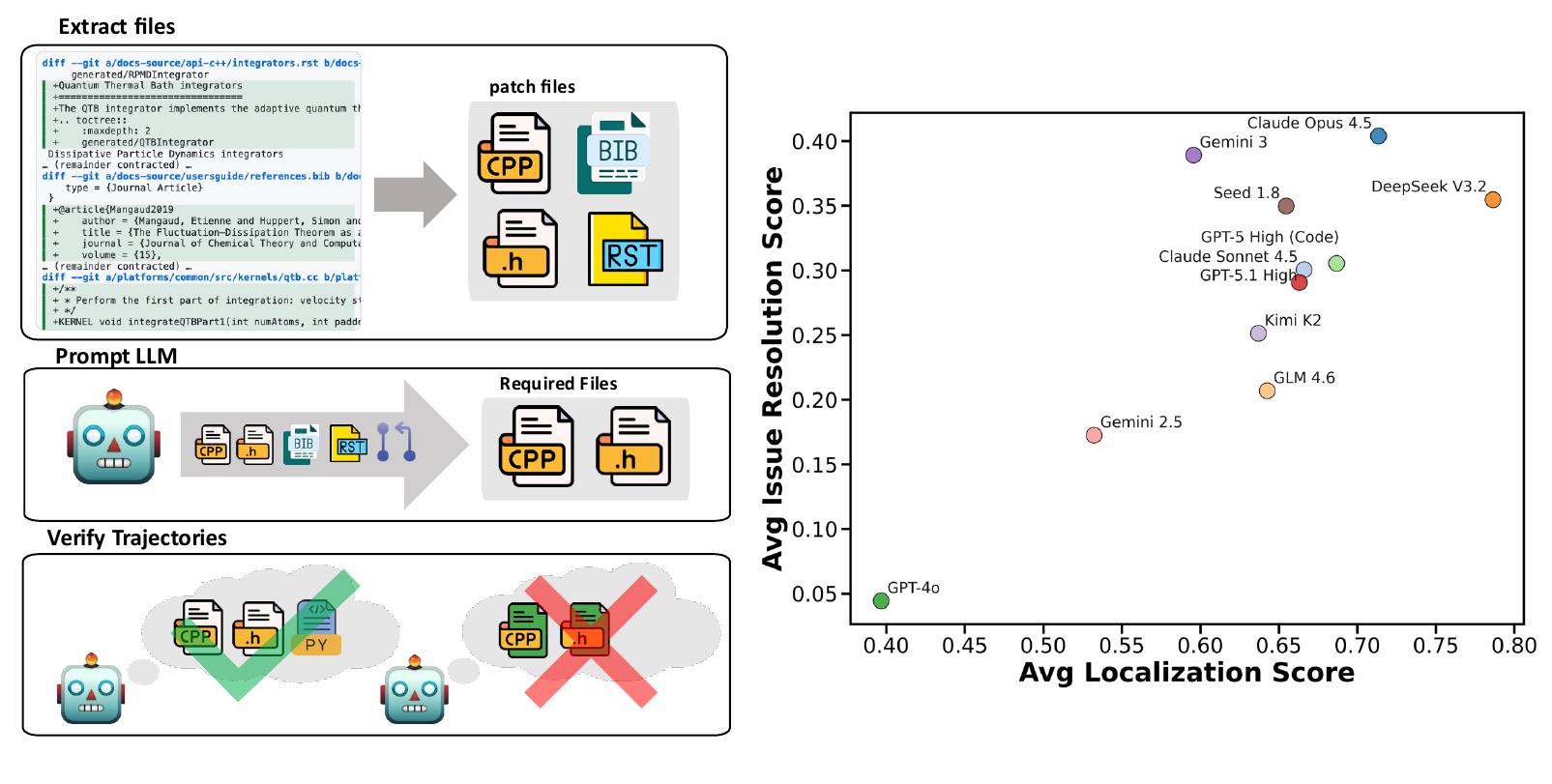}
    \caption{Localization pipeline. For each instance, we (1) extract all files modified by the gold patch; (2) prompt a SOTA LLM to identify which of these files are \textit{required} for test success based on the problem description, tests, and gold patch; and (3) score trajectories as localized if they issue any command involving required files (e.g., editing, viewing, or deleting).}
    \label{fig:model-local}
\end{figure}

As illustrated in the benchmark results, we observe a distinct stratification of model capabilities that suggests a shift in the primary bottlenecks of the field. While localizing complex repository-level issues remains an open challenge, top-performing models are increasingly proficient at identifying the relevant codebase coordinates. Models such as \textit{DeepSeek V3.2} and \textit{Claude Opus 4.5} achieve localization scores exceeding $0.70$, approaching a ceiling where the ``search'' phase is not a dominant failure mode. This indicates that the frontier of improvements ought to shift from retrieval-augmented discovery toward the higher-order reasoning required for functional synthesis.

\begin{figure}[h]
    \centering
    \includegraphics[width=0.3\linewidth]{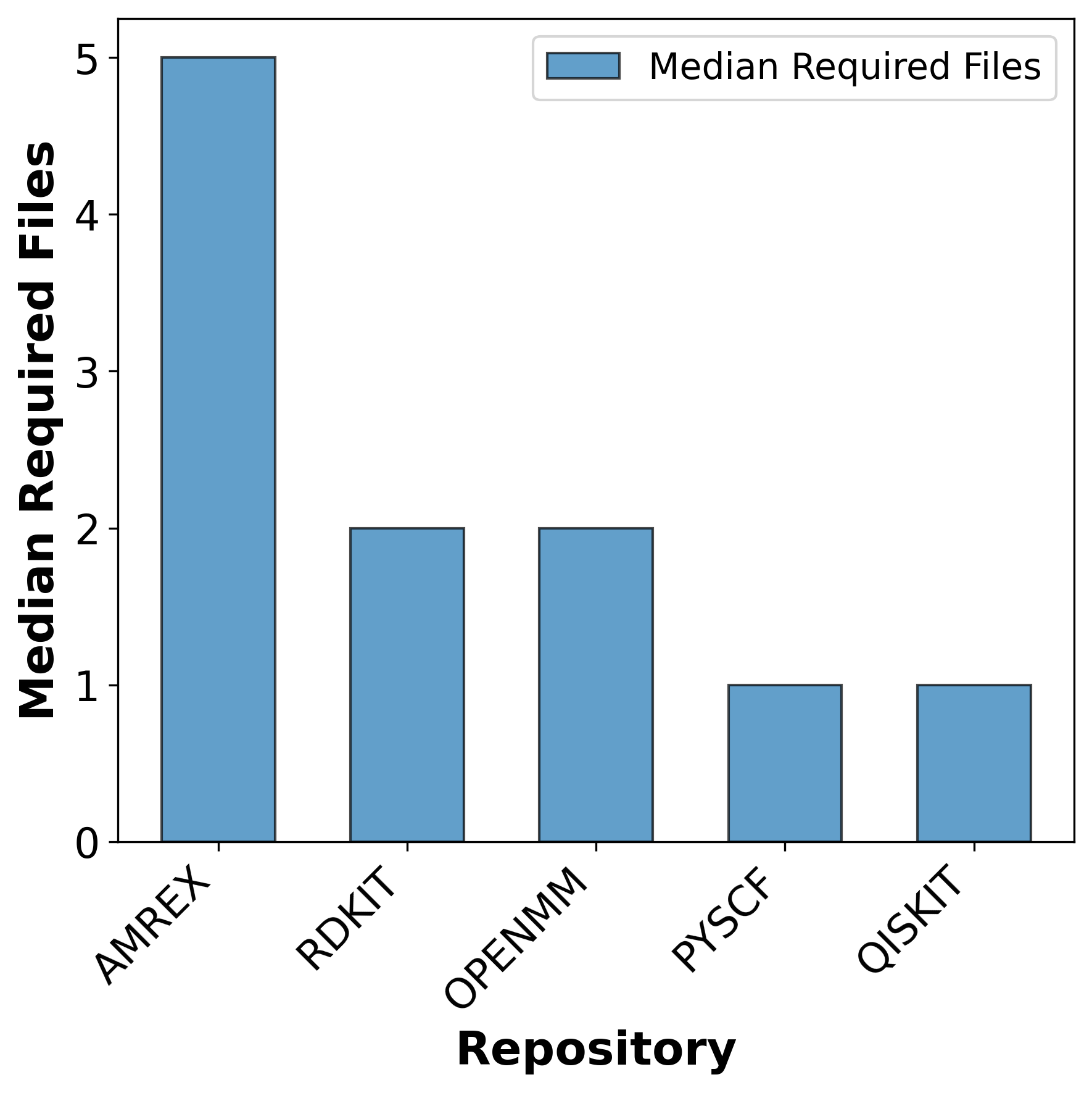}
    \caption{Median number of required files across repositories. Large number of required files implies strong cross-file reasoning required. Einstein-Toolkit is omitted from this graph since it's required files is always 1 by construction}
    \label{fig:model-loc-performance}
\end{figure}

In an idealized evaluation framework, implementation correctness is a strictly stronger condition than localization; a correct fix inherently implies the files were identified. In practice, however, binary localization metrics are inherently brittle. As the number of required files grows, the quality of the localization scores tend to degrade since models begin to identify valid, alternative fix locations which differ from the ground-truth ``required-file''. That said, Fig.~\ref{fig:model-loc-performance} indicates that most issues involve few required files, reducing ambiguity and preserving the practical utility of the localization metric.

 One stark exception to this is the AMReX repository, which exhibits the highest median number of required file at five. This increased dependency correlates with markedly lower resolve rates, highlighting a distinct failure mode in current frontier models: the lack of robust cross-file reasoning. Higher-order synthesis required to implement a consistent fix across five or more files remains a primary bottleneck, even among top-performing models.

\section{Discussions}

% \textcolor{red}
{
\subsection{Illustrative Case Studies.}
Aggregated success rates capture whether a model ultimately resolves a task, but they obscure the reasoning processes that lead to success or failure. To better interpret the results reported above, we examine several representative case studies that make explicit the kinds of scientific judgments required.}

In molecular simulation (Case Study~\ref{case:openmm-plasma-tp}), the task is not a just an issue to fix, it is missing an entire physical contribution: a self-energy term required for invariance under changes to PME parameters. This case study is particularly illustrative because resolution did not stop at correcting the underlying implementation, but required extending the existing testing infrastructure to distinguish genuine physical invariance from residual numerical error by systematically tightening grid resolution and tolerance parameters, reflecting domain-aware reasoning about Ewald truncation and discretization effects.

% \textbf{Promising Scientific Reasoning.} 
In quantum computing (Case Study~\ref{case:solovay-kitaev-phase}), 
{resolution begins by recognizing that, in Qiskit’s implementation, the recursive part of the Solovay–Kitaev algorithm operates on rotation matrices in SO(3) rather than on unitary matrices in SU(2). To apply the algorithm to decompose a unitary in SU(2), a mapping from SU(2) to SO(3) is therefore required. This mapping introduces a global-phase ambiguity, since two unitary matrices that differ only by a phase of $-1$ are mapped to the same element in SO(3). An LLM needs to understand both why the SU(2)-to-SO(3) conversion is necessary and what information may be lost in this process; however, LLM agents struggle with both aspects. Although it is also possible to implement the Solovay–Kitaev algorithm entirely within SU(2), doing so requires a deeper understanding of the algorithm and more extensive modifications to Qiskit’s code, which presents an equally significant challenge for LLM agents.
}

% Models that succeed do so \textcolor{red}{XXX}. 
In numerical relativity (Case Study~\ref{case:misner-wormhole-metric}), the overall correctness is determined by the numerical accuracy of the simulation using generated code versus the reference data (provided in the test). In the reasoning process, enforcing physical symmetries and understanding analytic limits are critical to accurately implementing codes. For instance, the example demonstrates the mistakes the model made in a) calculus limits at $n\rightarrow0$, b) missing absolute signs. On the other hand, once we realize these weaknesses and correct the code by changing just two lines, the results of the simulation become accurate. This shows that modern models are able to connect concise analytic constructions from the literature to their concrete numerical realization in large, unfamiliar HPC codebases, while maintaining stability and physical consistency.

\textbf{Failure Modes.}
Beyond the software engineering failure modes emphasized by SWE-bench–style evaluations \cite{yang2024sweagentagentcomputerinterfacesenable}, our benchmark exposes a distinct class of limitations that arise from scientific correctness rather than code repair. In these settings, failures are not only about locating files, satisfying tests, or repairing broken APIs, but about respecting domain-encoded invariants. Unlike the former, these constraints are fundamental to the underlying science. As a result, models often produce implementations that are locally plausible and numerically stable, but scientifically invalid.

Across repositories and scientific domains, we observe recurring errors that reflect gaps in mathematical precision, conservation laws, spatial reasoning, and adherence to established scientific conventions. These failures frequently manifest as subtle global inconsistencies: a correct-looking local fix that silently violates an invariant whose consequences only emerge downstream. The failure modes summarized in the table below illustrate that current frontier LLMs struggle most when correctness depends on real physical and mathematical laws, geometric reasoning, or community conventions. Together, these results suggest that scientific software repair poses challenges that are qualitatively different from existing SWE-centric benchmarks.

\begin{table}[H]
\centering
\small
\setlength{\tabcolsep}{7pt}
\renewcommand{\arraystretch}{1.15}
\begin{tabular}{p{0.26\linewidth} p{0.68\linewidth}}
\toprule
\textbf{Failure mode} & \textbf{Description} \\
\midrule
\textbf{Mathematics} &
Implements the right structure, but gets a crucial analytic detail wrong (e.g., missing a limit term or sign/absolute value in a series; Case~\ref{case:misner-wormhole-metric}. {Fail to notice the SU(2)-to-SO(3) conversion introduces phase ambiguity; Case ~\ref{case:solovay-kitaev-phase}}). \\

\textbf{Conservation Laws} &
Proposed fix breaks an exact invariant, or violates strict conservation conditions including of momentum, energy, or particle number. (e.g., electron-number conservation under smearing for odd $N$; Case~\ref{case:pyscf-rhf-smearing-odd-nelec}). \\

\textbf{Spatial Reasoning} &
Inability to reliably reason about global three-dimensional geometry and spatial constraints, resulting in implementations that satisfy local or axis-aligned criteria but violate true geometric guarantees (Case~\ref{case:openmm-solvent-geometry}). \\

\textbf{Scientific convention} &
Ignores or misapplies established scientific or codebase conventions that encode global invariants, causing subtle nonlocal inconsistencies despite locally reasonable fixes (Case~\ref{case:pyscf-bloch-phase-sign}). \\

\textbf{Scientific knowledge} &
Fails to distinguish scientifically meaningful constraints from representational artifacts, leading to fixes that stabilize behavior across input orderings or encodings while violating domain correctness (Case~\ref{case:rdkit-tautomerhashv2-order}). \\

\bottomrule
\end{tabular}
\label{tab:scientific-failure-modes}
\end{table}

% Beyond SWE based difficulties, models continue to struggle with XXX but with enforcing scientific invariants that are implicit, distributed, and rarely unit-tested in ordinary application code, but are critical in research-grade scientific software.

\subsection{Limitations}

% \textbf{\textsc{AInsteinBench}.}
Despite its scale and scientific grounding, \textsc{AInsteinBench} has several limitations.

First, \textit{test suites remain imperfect scientific oracles}. Although all instances are drawn from maintainer-accepted pull requests and undergo human verification, unit tests in scientific repositories frequently codify imperfect correctness conditions rather than fool-proof physical or mathematical invariants. As demonstrated by our false-positive and false-negative case studies, tests may be either too permissive, allowing scientifically invalid solutions, or overly prescriptive, rejecting alternative but valid implementations. While our curation pipeline mitigates the most egregious cases, completely eliminating such effects remains a major bottleneck in this pipeline.

Second, the benchmark is inherently \textit{biased by repository practices}. Scientific communities differ widely in testing culture, documentation quality, and software architecture. Repositories with well-maintained code, established and consistent infrastructure, and rich test harnesses are more amenable to PR-based benchmarking, while equally important scientific codes with sparse tests or monolithic designs are underrepresented. As a result, we admit that benchmark coverage is limited which adhere to software engineering norms.

Third, the benchmark is constrained to \textit{self-contained, lightweight execution environments}. Each task must run reliably inside a container on a small number of CPUs, and rely on off-the-shelf packages. As a result, we exclude pull requests whose validation depends on intensive computational resources (e.g., long-running simulations, large MPI jobs, GPU-heavy kernels), or runtime communication with external software systems. While such tasks are scientifically valuable and representative of real research workflows, incorporating them substantially increases the difficulty of the benchmark, with complexity compounding across environment reconstruction, data dependencies, and evaluation design. Crucially, this would also reduce reproducibility and accessibility for third parties, making independent verification and reuse significantly more challenging.

\subsection{Conclusion.}
% \textcolor{red}
{ \textsc{AInsteinBench} introduces the first large-scale benchmark that evaluates LLM agents inside real, production scientific software repositories. By grounding evaluation in executable environments, domain-specific failure modes, and maintainer-authored fixes, the benchmark moves beyond abstract scientific reasoning and conventional software-engineering tasks to assess whether models can function as genuine scientific computing agents. Our results highlight both encouraging progress and persistent gaps, particularly where complex spacial reasoning, hardware-level engineering, and cross-file reasoning are required. We hope this benchmark serves as a foundation for future work on reliable, domain-aware agents capable of assisting scientific software development at scale.}

\textsc{AInsteinBench} introduces the first large-scale benchmark that evaluates LLM agents within real, production scientific software repositories, grounded in executable environments, maintainer-authored fixes, and domain-specific correctness criteria. By design, the benchmark moves beyond abstract scientific reasoning and conventional software-engineering tasks, probing whether models can operate as genuine scientific computing agents capable of respecting physical invariants, mathematical structure, and established community conventions. Our results reveal both clear progress and persistent gaps, particularly when problems demand global spatial reasoning, hardware-level engineering insight, or tightly coupled cross-file updates.

At the same time, the benchmark surfaces encouraging evidence that frontier models can succeed on nontrivial scientific reasoning when the relevant structure is properly recognized. Successful trajectories include resolving subtle global-phase ambiguities in quantum compilation (Case Study~\ref{case:solovay-kitaev-phase}), enforcing analytic limits and physical symmetries in numerical relativity (Case Study~\ref{case:misner-wormhole-metric}), and identifying missing physical contributions in molecular simulation that require both implementation changes and test-suite extensions to validate invariance properties (Case Study~\ref{case:openmm-plasma-tp}). In these cases, models demonstrate the ability to connect compact theoretical insights to concrete, large-scale scientific codebases, going beyond surface-level fixes to reason about physical correctness.

Together, these results position \textsc{AInsteinBench} as both a stress test and a progress marker: it exposes scientific failure modes that lie outside the scope of existing SWE benchmarks, while also documenting that LLM agents are beginning to exhibit the kinds of domain-aware reasoning required to assist scientific software development at scale. We hope this benchmark serves as a foundation for future work on reliable, domain-aware agents capable of assisting scientific software development at scale.

\newpage

\clearpage
\newpage
\section{Contributions and Acknowledgments}
\label{sec:contributions}
The authors are listed in alphabetical order by their first names. Some names refer to the authors' internal aliases at the company.
\setlength{\parskip}{0pt} % 让段落之间没有额外空隙
\setlength{\itemsep}{0pt} % 如果用itemize
\setlength{\parsep}{0pt}  % 控制段落间距
\begin{multicols}{2}
\subsubsection*{Core Contributors}
%total: 49
Titouan Duston\\
Shuo Xin\\
Yang Sun\\
Daoguang Zan\\
Aoyan Li\\
Shulin Xin\\
Kai Shen \\
Yixiao Chen \\
Qiming Sun

\subsubsection*{Contributors}
%total: 136
Ge Zhang\\
Jiashuo Liu\\
Huan Zhou \\
Jingkai Liu \\
Zhichen Pu \\
Yuanheng Wang \\
Bo-Xuan Ge \\
% Bo-Xuan Ge$^1$ \\
Xin Tong \\
Fei Ye \\
% Zhi-Chao Zhao$^2$ \\
% Wen-Biao Han$^3$ \\
% Zhoujian Cao$^4$ \\
Zhi-Chao Zhao \\
Wen-Biao Han \\
Zhoujian Cao \\
Yueran Zhao \\
Weiluo Ren

\subsubsection*{Supportive Contributors}
Qingshen Long \\
Yuxiao Liu \\
Anni Huang \\
Yidi Du \\
Yuanyuan Rong \\
Jiahao Peng

\end{multicols}
% \subsubsection*{External Affiliations}

% $^1$ University of Chinese Academy of Sciences

% $^2$ China Agricultural University

% $^3$ Shanghai Astronomical Observatory

% $^4$ Beijing Normal University

\subsubsection*{Acknowledgments}
%total: 41
We sincerely thank Yuanyuan Rong, Jiahao Peng, Haili Wu, Shangshu Li, Huixiong Cao, Panchang Wei, and Xiaochuan She for their insightful discussions and continuous support. Their thoughtful feedback and expertise were instrumental in shaping the development, evaluation and exploration of future research directions for AInsteinBench.

\newpage

\appendix

\section{Problem Difficulty Grading Rubrics}
\label{sec:grading-rubrics}
Our expert team consists of 10 experts with PhD degree in the relevant field. We provide them the following rubrics for their annotation of problem difficulties. 

\textbf{Scientific Depth (1--5).}
Scientific depth measures the minimum level of domain-specific scientific knowledge required to correctly understand and resolve a task.

\begin{itemize}[
    leftmargin=*,
    itemsep=0.25em,
    parsep=0pt,
    topsep=0.25em,
    partopsep=0pt
]
\item \textbf{Score 1.} The task can be completed without domain-specific scientific knowledge; a non-expert can solve it by following explicit instructions or existing code patterns.
\item \textbf{Score 2.} The task requires basic undergraduate-level knowledge in the relevant scientific domain, such as standard definitions or textbook formulas.
\item \textbf{Score 3.} The task requires advanced undergraduate-level understanding, including reasoning about scientific assumptions, boundary conditions, or interactions among multiple domain concepts.
\item \textbf{Score 4.} The task requires graduate-level coursework knowledge or equivalent research exposure, such as familiarity with specialized algorithms or numerical methods used in the domain.
\item \textbf{Score 5.} The task requires research-level scientific expertise, including subtle theoretical considerations or domain-specific conventions typically acquired through active research.
\end{itemize}

\textbf{Engineering Difficulty (1--5).}
Engineering difficulty measures the software engineering complexity involved in implementing a correct solution.

\begin{itemize}[
    leftmargin=*,
    itemsep=0.25em,
    parsep=0pt,
    topsep=0.25em,
    partopsep=0pt
]
\item \textbf{Score 1.} Minimal engineering effort; the task can be completed without understanding internal APIs or implementation logic (e.g., simple copy--paste or parameter changes).
\item \textbf{Score 2.} Basic programming effort involving simple control flow or localized code modifications.
\item \textbf{Score 3.} Moderate implementation complexity, requiring nontrivial logic, advanced language features, or careful state management within a single component.
\item \textbf{Score 4.} Complex engineering effort involving coordination across multiple functions or files and understanding internal APIs.
\item \textbf{Score 5.} Large-scale engineering effort spanning multiple files, modules, or APIs, requiring architectural understanding and careful integration.
\end{itemize}

\section{Synthesized Question examples}
\label{app:ET_example}
\subsection{Prompt Template}

A question in our synthesized \verb|EinsteinToolkit| data can be formalized by the following statement. This is also the prompt template we send to LLM when evaluating their single-round performance (i.e., without using agent).
\begin{tcolorbox}[
    colback=green!5,
    colframe=green!40!black,
    left=2pt,right=2pt,top=2pt,bottom=2pt,
    boxrule=0.4pt,
    width=0.92\linewidth,
    enhanced jigsaw,
    breakable,
]
\scriptsize
\begin{Verbatim}[breaklines=true, breakanywhere=true]
You are an expert C/C++/Fortran developer working on EinsteinToolkit, a codebase for numerical relativity simulations.

Create C/C++/Fortran code for the file `{src_filename}` in thorn `{thorn_name}`.

## Thorn Information:
- Name: {thorn_name}
- Target file: {src_filename}

## Interface Definition in interface.ccl:
```
{interface}
```

## Schedule Definition in schedule.ccl:
```
{schedule}
```

## Parameters Definition in param.ccl:
```
{param}
```

## Configuration Definition in configuration.ccl:
```
{configuration}
```

{doc_context_section}

## Related Code Context:

{context}


## Instructions:
Generate only the complete C/C++/Fortran source code for `{src_filename}`. Include necessary headers, functions, and follow EinsteinToolkit conventions. 


IMPORTANT: Return ONLY the raw C/C++/Fortran source code. Do NOT use markdown code blocks (```c or ```). Do NOT include any explanations, comments outside the code, or formatting. Just return the plain C/C++/Fortran source code that can be directly saved to a .c/.cpp/.f file.

Code:

\end{Verbatim}
\end{tcolorbox}
where the \verb|{src_filename}|, \verb|{thorn_name}|, \verb|{interface}|, \verb|{schedule}|, \verb|{param}|, \verb|{configuration}|, \verb|{doc_context_section}|, \verb|{context}| are specific to each question. See the sample questions for an example.

\subsection{Sample Questions}

A question can be uniquely specified by \verb|{thorn_name}| and \verb|{src_filename}|, meaning that the corresponding src file was removed in the Thorn and the LLM is asked to generate the missing code. For LLM agent, we would provide a Docker container where the specified file is deleted and the agent is able to interact with the environment with shell command. For single-round generation, we provide the relevant context using the prompt template shown above. 

As an example, for the question where the LLM is tasked with implementing the Misner metric for wormhole spacetime, the corresponding \verb|{thorn_name}| is \verb|EinsteinInitialData/IDAnalyticBH|, and the corresponding \verb|{src_filename}| is \verb|Misner_standrad.c|. 

For LLM agent, we provide a Docker container where the relevant file is deleted. For single-round generation, we provided (to be replaced to the corresponding \verb|{...}| positions in the prompt template above).

\verb|{interface}|
\begin{tcolorbox}[
    colback=green!5,
    colframe=green!40!black,
    left=2pt,right=2pt,top=2pt,bottom=2pt,
    boxrule=0.4pt,
    width=0.92\linewidth,
    enhanced jigsaw,
    breakable,
]
\scriptsize
\begin{Verbatim}[breaklines=true, breakanywhere=true]
# Interface definition for thorn IDAnalyticBH
# $Header$

implements: idanalyticbh
inherits: ADMBase, StaticConformal, Grid

private:

#real misner_workspace TYPE = GF
#{
#  csch,
#  coth,
#  r1,
#  r2
#} "Workspace arrays for Misner_standard"
\end{Verbatim}
\end{tcolorbox}

\verb|{schedule}| (the complete file is too long and we omitted multiple parts in "...")
\begin{tcolorbox}[
    colback=green!5,
    colframe=green!40!black,
    left=2pt,right=2pt,top=2pt,bottom=2pt,
    boxrule=0.4pt,
    width=0.92\linewidth,
    enhanced jigsaw,
    breakable,
]
\scriptsize
\begin{Verbatim}[breaklines=true, breakanywhere=true]
# Schedule definitions for thorn IDAnalyticBH
# $Header$

if (CCTK_Equals(initial_data,"schwarzschild") ||
    CCTK_Equals(initial_data,"kerr") ||
    CCTK_Equals(initial_data,"bl_bh") ||
    CCTK_Equals(initial_data,"misner_bh") ||
    CCTK_Equals(initial_data,"multiple_misner"))
{
  schedule IDAnalyticBH_ParamChecker at CCTK_PARAMCHECK
  {
    LANG: C
    OPTIONS: global
  } "Construct parameters for analytic black hole solutions"
}

if (CCTK_Equals(initial_data,"schwarzschild")) 
{ 
	
   schedule Schwarzschild in ADMBase_InitialData
   {
     STORAGE: confac[1]
     LANG: C
   } "Construct initial data for a single Schwarzschild black hole"
}
... more schedules omitted...
else if (CCTK_Equals(initial_data,"multiple_misner_bh")) 
{ 
   schedule Misner_multiple in ADMBase_InitialData
   {
     STORAGE: confac[1]
     LANG: C
   } "Construct initial data for multiple Misner black holes"

}
\end{Verbatim}
\end{tcolorbox}

\verb|{param}| (the complete file is too long and we omitted multiple parts in "...")
\begin{tcolorbox}[
    colback=green!5,
    colframe=green!40!black,
    left=2pt,right=2pt,top=2pt,bottom=2pt,
    boxrule=0.4pt,
    width=0.92\linewidth,
    enhanced jigsaw,
    breakable,
]
\scriptsize
\begin{Verbatim}[breaklines=true, breakanywhere=true]
# Parameter definitions for thorn IDAnalyticBH
# $Header$

private:

# Schwarzschild parameters
# ------------------------
REAL mass "Mass of black hole"
{
: :: "Not sure if it can be negative or not"
} 2.0

# Kerr parameters
# ------------------------
REAL a_Kerr "Angular momentum parameter of black hole"
{
-1:1 :: "Between +1 and -1"
} 0.1

# Multiple Misner Parameters
# --------------------------
REAL mu "Misner mu value"
{
0: :: "Non-negative"
} 1.2

...more parameters omitted...

USES KEYWORD metric_type 

shares: StaticConformal

USES KEYWORD conformal_storage

\end{Verbatim}
\end{tcolorbox}

\verb|{configuration}|
\begin{tcolorbox}[
    colback=green!5,
    colframe=green!40!black,
    left=2pt,right=2pt,top=2pt,bottom=2pt,
    boxrule=0.4pt,
    width=0.92\linewidth,
    enhanced jigsaw,
    breakable,
]
\scriptsize
\begin{Verbatim}[breaklines=true, breakanywhere=true]
# configuration is empty for this Thorn
\end{Verbatim}
\end{tcolorbox}

\verb|{doc_context_section}| (the complete file is too long and we omitted multiple parts in "...")
\begin{tcolorbox}[
    colback=green!5,
    colframe=green!40!black,
    left=2pt,right=2pt,top=2pt,bottom=2pt,
    boxrule=0.4pt,
    width=0.92\linewidth,
    enhanced jigsaw,
    breakable,
]
\scriptsize
\begin{Verbatim}[breaklines=true, breakanywhere=true]

## README (README):
```
Cactus Code Thorn IDAnalyticBH
...more readme file content omitted...
This thorn calculates analytic initial data for the Einstein grid functions
(lapse, shift, metric, curv) for various black hole spacetimes:

	Schwarzschild black hole
 	2 Misner black holes placed on the z-axis
	n Misner black holes placed in a circle in the x-y plane
	Brill-Lindquist data
```

## Documentation (.tex files):
Files: documentation.tex

```tex
=== documentation.tex ===
% /*@@
%   @file      documentation.tex
%   @date      28 April 2002
%   @author    Denis Pollney

...more latex documentation omitted...

\subsection{Two-throat Misner data}

The \texttt{misner\_bh} initial data generates a metric of the form
\begin{equation}
  ds^2 = -dt^2 + \psi^4 (dx^2 + dy^2 + dz^2),
\end{equation}
where the conformal factor $\psi$ is given by
\begin{equation}
  \psi = \sum^N_{n=-N}
  \frac{1}{\sinh(\mu_0 n)}
  \frac{1}{\sqrt{x^2 + y^2 + (z + \coth(\mu_0 n))^2}}.
\end{equation}
The extrinsic curvature for the Misner data is zero.

...more latex documentation omitted...

    \textit{Interaction Energy in Geometrostatics}
    Phys. Rev., \textbf{131}, 471--476.
\end{thebibliography}

% Do not delete next line
% END CACTUS THORNGUIDE
\end{document}
```
\end{Verbatim}
\end{tcolorbox}

\verb|{context}| (the complete file is too long and we omitted multiple parts in "...")
\begin{tcolorbox}[
    colback=green!5,
    colframe=green!40!black,
    left=2pt,right=2pt,top=2pt,bottom=2pt,
    boxrule=0.4pt,
    width=0.92\linewidth,
    enhanced jigsaw,
    breakable,
]
\scriptsize
\begin{Verbatim}[breaklines=true, breakanywhere=true]

Kerr.c:
```
 /*@@
   @file      Kerr.c

...more codes omitted...

  return;
}

```

IDAnalyticBH.h:
```
 /*@@
   @file      IDAnalyticBH.h

...more code omitted...

/* misc.c */
void IDAnalyticBH_zero_CCTK_REAL_array(int N, CCTK_REAL A[]);

```

ParamChecker.c:
```
 /*@@
   @file      ParamChecker.F

...more code omitted...

    CCTK_INFO("Implements non-conformal metric");
    CCTK_INFO("  (Not usually a good idea!)");
  }        

}

```

make.code.defn:
```
# Main make.code.defn file for thorn IDAnalyticBH
# $Header$

# Source files in this directory
SRCS = 	ParamChecker.c\
	Schwarzschild.c\
	Kerr.c\
	BrillLindquist.c\
        Misner_standard.c\
	Misner_multiple.c\
	Misner_points.c \
        misc.c

# Subdirectories containing source files
SUBDIRS = 
```

Misner_multiple.c:
```
 /*@@
   @file      Misner_multiple.F

...more code omitted...

  IDAnalyticBH_zero_CCTK_REAL_array(npoints, kzz);

}

```

misc.c:
```
 /*@@
   @file      misc.c

...more code omitted...

  for (i = 0 ; i < N ; ++i)
  {
    A[i] = 0.0;
  }
}

```

BrillLindquist.c:
```
 /*@@
   @file      BrillLindquist.F

...more code omitted...
  IDAnalyticBH_zero_CCTK_REAL_array(npoints, kzz);
}

```

Misner_points.c:
```
 /*@@
   @file      Misner_points.c

...more code omitted...

  res += b->mass/sqrt(
      (x - b->x)*(x - b->x)+
      (y - b->y)*(y - b->y)+
      z*z
    );
  return res;
}

```

Schwarzschild.c:
```

 /*@@
   @file      Schwarzschild.c

...more code omitted...

  IDAnalyticBH_zero_CCTK_REAL_array(npoints, kzz);
}
```
\end{Verbatim}
\end{tcolorbox}

\section{Case Studies}

\begin{casebox}[top=2pt,bottom=2pt,left=4pt,right=4pt,breakable]{
\vspace{-0.6\baselineskip}

\subsection{Neutralizing Plasma Correction in OpenMM Ewald/PME}}\refstepcounter{case}\label{case:openmm-plasma-tp}

% \noindent\makebox[0pt][l]{\rule{\dimexpr\linewidth\relax}{0.8pt}}\par

\scriptsize   % 比 \small 再小一档，整体更紧凑

\textbf{Pull Request Description.}

\medskip

**Problem:**
The energy calculations for systems using the AMOEBA and HIPPO force fields do not account for the correction due to the neutralizing plasma in periodic boundary conditions. This omission can lead to inaccuracies in the computed potential energy for systems with a net nonzero charge, particularly when using Particle Mesh Ewald (PME) for long-range electrostatics.

**Expected Behavior:**
The energy calculations should include a correction term for the neutralizing plasma when PME is used. This ensures that the potential energy of systems with a net charge is accurately computed, consistent with the physical behavior of periodic systems. The energy should also remain consistent across different PME parameters, such as the alpha value, for systems with nonzero net charge.

% OpenMM’s AMOEBA and HIPPO electrostatics use Ewald / PME with periodic boundary conditions.  
% In PR~\#4920, the maintainer reports that systems with nonzero net charge show a strong, unphysical dependence of the potential energy on the Ewald splitting parameter \(\alpha\), because the standard neutralizing plasma self-energy term is missing from the AMOEBA/HIPPO PME kernels. The intended behavior is that when PME is enabled, the energy should include the neutralizing plasma correction
% \[
% E_{\text{plasma}} = - \frac{\pi}{2 V \alpha^2} Q^2 \times \frac{1}{4\pi\epsilon_0},
% \]
% so that charged systems have \(\alpha\)-invariant energies (up to small numerical PME errors) and are consistent across different PME parameter choices.

\medskip

\textbf{What did the model do?}

The model first explored the OpenMM repository, locating the AMOEBA/HIPPO reference and common kernels and confirming the Python wrappers for `AmoebaMultipoleForce`. It then constructed a focused Python reproduction: a single charged particle in a periodic box with AMOEBA multipoles and PME enabled, comparing energies at \(\alpha=3.0\) and \(\alpha=4.0\); the large \(\sim 1.3\ \text{kJ/mol}\) discrepancy confirmed strong, unphysical \(\alpha\)-dependence.  
Using Ewald theory, the model hypothesized that the missing neutralizing plasma term was the culprit, derived the correction \(E_{\text{plasma}} = -\frac{\pi}{2V\alpha^2}Q^2(1/4\pi\epsilon_0)\), and mapped each symbol to OpenMM’s internal quantities: total charge from multipole parameters, volume from periodic box vectors, and \(1/4\pi\epsilon_0\) from the existing $\frac{`EPSILON0`}{ `_electric`}$ constants. It then implemented the correction in the AMOEBA common PME kernel (and analogously for HIPPO), ensuring the term is applied only when PME is active and accumulating total charge once during parameter initialization / updates.

\begin{tcolorbox}[
    colback=green!5,
    colframe=green!40!black,
    left=2pt,right=2pt,top=2pt,bottom=2pt,
    boxrule=0.4pt,
    width=0.92\linewidth,
    enhanced jigsaw,
    breakable,
]
\scriptsize
\begin{Verbatim}[breaklines=true, breakanywhere=true]
 // AMOEBA multipoles: accumulate total charge during initialization.
 void CommonCalcAmoebaMultipoleForceKernel::initialize(const System& system, const AmoebaMultipoleForce& force) {
     ...
     totalCharge = 0.0;
     for (int i = 0; i < numMultipoles; i++) {
         double charge, thole, damping, polarity;
         int axisType, atomX, atomY, atomZ;
         vector<double> dipole, quadrupole;
         force.getMultipoleParameters(i, charge, dipole, quadrupole, axisType,
                                      atomZ, atomX, atomY, thole, damping, polarity);
         totalCharge += charge;
         ...
     }
     ...
 }

 // In the PME execution, add the neutralizing plasma correction.
 double CommonCalcAmoebaMultipoleForceKernel::execute(ContextImpl& context, bool includeForces,
                                                      bool includeEnergy) {
     ...
     cc.getPosq().copyTo(lastPositions);
     multipolesAreValid = true;

     if (usePME) {
         Vec3 a, b, c;
         cc.getPeriodicBoxVectors(a, b, c);
         double volume = a[0]*b[1]*c[2];   // orthorhombic box
         return -totalCharge*totalCharge/(8*EPSILON0*volume*pmeAlpha*pmeAlpha);
     }
     else
         return 0.0;
 }
\end{Verbatim}
\end{tcolorbox}

\textbf{Why is this a strong scientific fix?}

After inserting the correction, the model rebuilt OpenMM and reran the reproduction: the \(\alpha\)-dependence shrank from \(\sim 1.3\ \text{kJ/mol}\) to \(\sim 10^{-3}\ \text{kJ/mol}\), and further tuning PME grid dimensions and error tolerances reduced the residual difference to \(\sim 2\times 10^{-5}\ \text{kJ/mol}\). The model also extended the fix to HIPPO by summing `coreCharge + valenceCharge` per particle, and verified that existing C++ tests for AMOEBA and HIPPO reference implementations still passed, indicating no regression in other electrostatics behavior.  
Crucially, the patch respects unit consistency and physical constants, correctly handles periodic box volume, and ensures that the correction is applied only in the PME pathway, aligning with the analytical Ewald limit for charged periodic systems.

\medskip

\textbf{Scientific intent vs. numerical artifacts.}

The original tests did not explicitly enforce \(\alpha\)-invariance for net-charged AMOEBA/HIPPO systems, so the missing plasma term could have remained hidden. The final verification distinguished true physical invariance from residual numerical PME errors by tightening grid and tolerance parameters, demonstrating domain-aware reasoning about Ewald truncation and discretization effects.

\scriptsize

\end{casebox}

\begin{casebox}[top=2pt,bottom=2pt,left=4pt,right=4pt,breakable]{
\vspace{-0.6\baselineskip}
\subsection{Unexpected Global Phase in Solovay--Kitaev Decomposition}}\refstepcounter{case}\label{case:solovay-kitaev-phase}
\scriptsize

{\normalsize\textbf{Pull Request Description}}

\medskip
\textbf{Environment}
\begin{itemize}[nosep,leftmargin=*]
    \item Qiskit Terra version: \texttt{0.23.1}
    \item Python version: \texttt{3.10.6}
    \item Operating system: \textbf{Ubuntu Linux}
\end{itemize}

\medskip
\textbf{What is happening?}\\
The global phase of some circuits generated by \texttt{SolovayKitaev} seems wrong.

\medskip
\textbf{How can we reproduce the issue?}

\begin{lstlisting}[mathescape=true,
    basicstyle=\ttfamily\scriptsize,
    aboveskip=1pt,
    belowskip=1pt,
    lineskip=-1pt,
    columns=fullflexible
]
>>> from qiskit.transpiler.passes.synthesis import SolovayKitaev
>>> from qiskit.quantum_info import Operator
>>> from qiskit import QuantumCircuit, Aer, transpile, assemble
>>> c = QuantumCircuit(1)
>>> c.y(0)
>>> skd = SolovayKitaev(recursion_degree=3)
>>> dc = skd(c)
>>> dc.draw() # global phase: $\frac{3\pi}{4}$
\end{lstlisting}

\medskip
\textbf{What should happen?}\\
Given \texttt{Operator(dc).data} is approximately equal to
\(
\begin{bmatrix}
0 & i \\
-i & 0
\end{bmatrix}
\),
which equals \(-\texttt{Operator(c).data}\),
the global phase of \texttt{dc} should be \(\pi\) (The correct global phase should be $-\pi/4$ instead of $\pi$) rather than \(3\pi/4\).

\medskip
{\normalsize\textbf{Core Fix}}
{
In Qiskit's implementation of the Solovay--Kitaev algorithm, the decomposition is performed in SO(3) rather than in SU(2). The conversion map from SU(2) to SO(3) is two-to-one and therefore introduces
}
a possible global phase flip of $\pm 1$.
The fix adds an explicit phase-alignment check and adjusts the resulting
operator's global phase angle by $\pi$ whenever the negated approximation matches the target better.

\begin{tcolorbox}[
    colback=green!5,
    colframe=green!40!black,
    left=2pt,right=2pt,top=2pt,bottom=2pt,
    boxrule=0.4pt,
    width=0.92\linewidth,
    enhanced jigsaw,
    breakable,
]
\scriptsize
\begin{Verbatim}[breaklines=true, breakanywhere=true]
+ def _should_adjust_phase(computed: np.ndarray, target: np.ndarray) -> bool:
+     """
+     The implemented SolovayKitaevDecomposition has a global phase
+     uncertainty of +-1, due to approximating not the original SU(2)
+     matrix but its projection onto SO(3). This function returns True
+     if the global phase of the computed approximation should be
+     adjusted (by adding pi) to better match the target.
+     """
+     return np.linalg.norm(-computed - target) < np.linalg.norm(computed - target)
\end{Verbatim}
\end{tcolorbox}

\medskip

\begin{tcolorbox}[
    colback=green!5,
    colframe=green!40!black,
    left=2pt,right=2pt,top=2pt,bottom=2pt,
    boxrule=0.4pt,
    width=0.92\linewidth,
    enhanced jigsaw,
    breakable,
]
\scriptsize
\begin{Verbatim}[breaklines=true, breakanywhere=true]
+ # adjust to the correct SU(2) phase
+ adjust_phase = (
+     np.pi if _should_adjust_phase(decomposition._to_u2(), gate_matrix_su2)
+     else 0.0
+ )
\end{Verbatim}
\end{tcolorbox}

\tcblower
{\footnotesize
\textbf{Summary.}
{
The Solovay--Kitaev algorithm efficiently approximates an arbitrary single-qubit gate in SU(2) using gates from a fixed, finite gate set, typically $\{T, T^\dagger, H\}$. In Qiskit’s implementation, this is achieved by first converting a unitary matrix in SU(2) to a rotation matrix in SO(3), and then performing the core decomposition in SO(3).
}
This example illustrates a scientifically meaningful bug arising from the fact that the homomorphism from SU(2) to SO(3) is two-to-one. As a result, the global phase has to be stored separately in order to reconstruct the original  SU(2) unitary from its SO(3) representation. In the Core Fix, such phase ambiguity is resolved through an extra phase adjust.} It demonstrates the need for domain-aware debugging beyond standard software-engineering fixes.

\end{casebox}

\begin{casebox}[top=2pt,bottom=2pt,left=4pt,right=4pt,breakable]{
\vspace{-0.6\baselineskip}
\subsection{Implement Misner metric for wormhole spacetime}}\refstepcounter{case}\label{case:misner-wormhole-metric}
    \scriptsize
    
    {\normalsize\textbf{The Task}}
    
    The LLM was asked to complete the \verb|Misner_standard.c| for setting up initial data of wormhole spacetime in the \verb|EinsteinToolkit| framework. The \texttt{IDAnalyticBH} thorn provides analytically-specified initial data for black hole evolutions, including the Misner solution~\cite{misner1960wormhole}, an early attempt to represent wormholes as initial data for Einstein's equations.
    
    \medskip
    {\normalsize\textbf{Physical Context}}\\
    Misner's 1960 solution describes two massive objects instantaneously at rest whose ``throats'' are connected through a wormhole topology. Unlike simple superposition, this is an exact solution to Einstein's constraint equations. The spacetime can be represented, in a proporly chosen coordinate, by a time-symmetric (vanishing extrinsic curvature $K_{ij}=0$) and conformally flat metric
    \[
    ds^2 = -dt^2 + \psi^4 (dx^2 + dy^2 + dz^2).
    \]
    The physics is encoded entirely in the conformal factor $\psi$, which solves the Hamiltonian constraint (reduces to $\nabla^2\psi = 0$ for this case). Misner's ingenious method of images constructs $\psi$ by summing contributions from image sources at locations determined by the wormhole topology.
    
    The \texttt{IDAnalyticBH} thorn implementation inherits metric grid-functions from \texttt{ADMBase} and conformal factor variables from \texttt{StaticConformal}. The code must: (1)~iterate over all spatial grid points, (2)~at each point $(x,y,z)$, sum the series for $\psi$ up to truncation limit \texttt{nmax}, (3)~set the conformal metric components to $g_{ij} = \delta_{ij}$ and extrinsic curvature to zero, and (4)~when \texttt{metric\_type="static conformal"}, compute and store analytical derivatives $\psi_{,i}$ and possibly $\psi_{,ij}$. The parameter $\mu$ (denoted \texttt{mu} in code) controls the throat of the wormhole.
    
    \medskip
    {\normalsize\textbf{Domain Knowledge}}\\
    Implementing this requires understanding:
    
    \begin{enumerate}[nosep,leftmargin=*]
        \item \textit{Initial value formulation}: The Einstein equations split into constraint equations (to be satisfied on an initial slice) and evolution equations. For time-symmetric data with $K_{ij}=0$, the constraint reduces to ${}^{(3)}R=0$ on the spatial slice.
        \item \textit{Conformal method}: Writing $g_{ij} = \psi^4 \delta_{ij}$ with flat conformal metric $\delta_{ij}$ reduces the constraint to $\nabla^2\psi = 0$ (Laplace's equation), enabling the method of images.
        \item \textit{Method of images}: The wormhole has periodicity in a toroidal coordinate $\mu$ with period $2\mu_0$. Image sources appear at $z_n = \coth(\mu_0 n)$ for all integers $n$, each contributing mass $\propto |\sinh(\mu_0 n)|^{-1}$.
        \item \textit{Coordinate transformation}: The natural toroidal coordinates $(\mu, \theta, \varphi)$ must be converted to Cartesian $(x,y,z)$ for numerical simulation. Each image source at $z_n$ contributes $\text{csch}(\mu_0 n) / r_n$ where $r_n = \sqrt{x^2 + y^2 + (z - z_n)^2}$.
    \end{enumerate}
    
    \medskip
    {\normalsize\textbf{Reference Answer}}\\
    The key computational task is evaluating the conformal factor $\psi$ at each grid point. From Misner's derivation, going to Cartesian coordinates, the conformal factor is:
    \[
      \psi = \sum^{\infty}_{n=-\infty}
      \frac{1}{|\sinh(\mu_0 n)|}
      \frac{1}{\sqrt{x^2 + y^2 + (z + \coth(\mu_0 n))^2}}.
    \]
    The sum extends over all integers $n$, with each term representing a gravitational image source. The $n=0$ term is indeterminate ($0/0$); applying L'Hôpital's rule yields the constant 1.0. The absolute value $|\sinh(\mu_0 n)|$ in the denominator is coming from pulling a complete square out of the root. The reference implementation exploits the symmetry $n \leftrightarrow -n$ and precomputes coefficients for $n \ge 1$:
    
\begin{tcolorbox}[
    colback=green!5,
    colframe=green!40!black,
    left=2pt,right=2pt,top=2pt,bottom=2pt,
    boxrule=0.4pt,
    width=0.92\linewidth,
    enhanced jigsaw,
    breakable,
]
\scriptsize
\begin{Verbatim}[breaklines=true, breakanywhere=true]
    for(n = 1; n <= nmax; n++) {
      csch[n] = 1.0 / sinh(mu*n);  // Always positive
      coth[n] = 1.0 / tanh(mu*n);
      adm_mass += 4.0 * csch[n];
    }
    \end{Verbatim} 
    \end{tcolorbox}
    
    Then for each grid point, it initializes \texttt{psi[i] = 1.0} (the $n=0$ contribution) and sums paired terms:
    
\begin{tcolorbox}[
    colback=green!5,
    colframe=green!40!black,
    left=2pt,right=2pt,top=2pt,bottom=2pt,
    boxrule=0.4pt,
    width=0.92\linewidth,
    enhanced jigsaw,
    breakable,
]
\scriptsize
\begin{Verbatim}[breaklines=true, breakanywhere=true]
    psi[i] = 1.0;  // n=0 limit
    for(n = 1; n <= nmax; n++) {
      inv_r1 = 1.0 / sqrt(xy_squared + (z[i]+coth[n])$^2$);
      inv_r2 = 1.0 / sqrt(xy_squared + (z[i]-coth[n])$^2$);
      psi[i] += csch[n]*(inv_r1 + inv_r2);  // Pairs +n and -n
    }
    \end{Verbatim}
    \end{tcolorbox}
    
    This pairing of sources at $\pm n$ automatically sums from $-N_{\rm max}$ to $+N_{\rm max}$ even though the code only explicitly iterate over $1 \rightarrow N_{\rm max}$

\medskip
{\normalsize\textbf{Failure patterns}}\\
The generated code failed numerical tests (\texttt{grr\_max: 2.26} vs.\ benchmark \texttt{80.18}) despite correct overall structure. It chose to iterate explicitly from $-N$ to $+N$ rather than exploiting symmetry:

\begin{tcolorbox}[
    colback=green!5,
    colframe=green!40!black,
    left=2pt,right=2pt,top=2pt,bottom=2pt,
    boxrule=0.4pt,
    width=0.92\linewidth,
    enhanced jigsaw,
    breakable,
]
\scriptsize
\begin{Verbatim}[breaklines=true, breakanywhere=true]
psi_val = 0.0;  // Error: n=0 limit is 1.0, not 0
for (n = -nmax; n <= nmax; n++) {
  if (n == 0) continue;
  csch_val = 1.0 / sinh(mu_val * n);  // Negative for n < 0
  r1 = sqrt(xx*xx + yy*yy + (zz + coth_val)*(zz + coth_val));
  psi_val += csch_val / r1;  // Error: missing fabs()
}
\end{Verbatim}
\end{tcolorbox}

Two physical errors: (1)~The $n=0$ term $\lim_{n\to 0} [\sinh(\mu n)]^{-1} / \sqrt{x^2+y^2+(z+\coth(\mu n))^2}$ evaluates to 1.0, seen explicitly by applying L'Hôpital's rule, not 0. (2)~Since $\sinh(\mu n)$ is negative for $n<0$, omitting \texttt{fabs()} causes negative mass contributions, violating physics, though the original absolute sign is coming from a calculus trick pulling a complete square out of the root. The reference code avoids both by pairing $\pm n$ terms and only sum over positive $n$. The fix (that we can do) is simple:

\begin{tcolorbox}[
    colback=green!5,
    colframe=green!40!black,
    left=2pt,right=2pt,top=2pt,bottom=2pt,
    boxrule=0.4pt,
    width=0.92\linewidth,
    enhanced jigsaw,
    breakable,
]
\scriptsize
\begin{Verbatim}[breaklines=true, breakanywhere=true]
-   psi_val = 0.0;
+   psi_val = 1.0;  // n=0 contribution
    ...
-   psi_val += csch_val / r1;
+   psi_val += fabs(csch_val) / r1;
\end{Verbatim}
\end{tcolorbox}

After correction: \texttt{grr\_max} 2.26 $\to$ 80.18, \texttt{grr\_min} $<$0.001 $\to$ 1.69 (matching benchmark).

\medskip
{
{\normalsize\textbf{Significance}}\\
This failure demonstrates the gap between general coding ability and domain expertise. The LLM successfully navigated the Cactus framework (understanding \texttt{ADMBase} and \texttt{StaticConformal} conventions, computing derivatives conditionally, handling grid iteration), and translated mathematics to code competently. However, it missed physics-driven implementation choices: the reference code's paired summation is not arbitrary style but encodes the physical symmetry $n \leftrightarrow -n$, automatically ensuring positive mass contributions. The LLM's explicit $-N$ to $+N$ iteration is mathematically equivalent but fails to recognize that (1)~the absolute value in the formula is not decorative but enforces physical positivity, requiring \texttt{fabs()} when symmetry isn't exploited, and (2)~the $n=0$ singularity requires limiting analysis yielding 1.0, not naive exclusion from a zero-initialized sum. These are not programming errors detectable by compilers or unit tests. Only domain knowledge and numerical validation against known solutions reveal them.
}
\end{casebox}

\begin{casebox}[top=2pt,bottom=2pt,left=4pt,right=4pt,breakable]{
\vspace{-0.6\baselineskip}
\subsection{Implement spin-separated 1--4 RDM APIs in PySCF FCI}}\refstepcounter{case}\label{case:pyscf-fci-rdm1234s} \scriptsize

{\normalsize\textbf{The Task}}\\
The agent was asked to extend PySCF's Full Configuration Interaction (FCI) solver with new APIs that produce reduced density matrices (RDMs) up to fourth order, including spin-resolved blocks. Concretely, two feature functions were to be added to \texttt{pyscf/fci/direct\_spin1.py}:
\begin{enumerate}[nosep,leftmargin=*]
    \item \texttt{make\_rdm1234}: return spin-traced 1-, 2-, 3-, and 4-RDMs following PySCF's FCI conventions.
    \item \texttt{make\_rdm1234s}: return spin-separated components, including 3-RDM blocks \texttt{(aaa, aab, abb, bbb)} and 4-RDM blocks \texttt{(aaaa, aaab, aabb, abbb, bbbb)}.
\end{enumerate}
The implementation needed to respect existing operator-order conventions and a \texttt{reorder} flag that maps internal kernel outputs into the user-facing canonical ordering. The accompanying tests assert that spin-separated blocks reconstruct the spin-traced tensors when summed with the required index transpositions.

\medskip
{\normalsize\textbf{Minimal API Context}}\\
The problem statement provided typed function signatures:
\vspace{0.2cm}
\begin{lstlisting}[basicstyle=\ttfamily\scriptsize,aboveskip=1pt,belowskip=1pt,columns=fullflexible]
def make_rdm1234(fcivec, norb, nelec, link_index, reorder) -> numpy.ndarray: ...
def make_rdm1234s(fcivec, norb, nelec, link_index, reorder) -> numpy.ndarray: ...
\end{lstlisting}

\vspace{0.2cm}

Despite the brevity, the naming and parameter patterns encoded strong affordances: \texttt{make\_rdm1234} clearly extends the existing \texttt{make\_rdm123} family in \texttt{direct\_spin1.py}, while the suffix \texttt{s} matches established spin-separated routines (\texttt{make\_rdm1s}, \texttt{make\_rdm12s}, \texttt{make\_rdm123s}). The shared argument schema \texttt{(fcivec, norb, nelec, link\_index, reorder)} aligns with PySCF's FCI interfaces and effectively points to the correct file and helper kernels.

\medskip
{\normalsize\textbf{Physical Context}}\\
Reduced density matrices encode $k$-body correlations of an $N$-electron wavefunction $|\Psi\rangle$ via expectation values of strings of creation/annihilation operators. PySCF adopts a consistent convention in which higher-order tensors follow
\[
\mathrm{dm}[p,q,r,s,\dots] \;=\; \langle p^\dagger r^\dagger \cdots s\,q\rangle,
\]
with the 1-RDM matching the mean-field convention. Spin separation decomposes the RDM into blocks based on the spin labels ($\alpha$ or $\beta$) of the involved orbital indices. For example, the 3-RDM decomposes into \texttt{aaa, aab, abb, bbb} blocks, and the 4-RDM into \texttt{aaaa, aaab, aabb, abbb, bbbb}. These spin-resolved objects enable diagnostics and development work where the distribution of spin among correlated electrons matters.

% \medskip
% {\normalsize\textbf{Documentation Discovered}}\\
% Guided by the minimal API spec, the agent traced the relevant implementation patterns in:
% \begin{itemize}[nosep,leftmargin=*]
%     \item \texttt{pyscf/fci/direct\_spin1.py}: existing \texttt{make\_rdm123} and \texttt{make\_rdm123s} provide a template for extending to 4-RDMs and for the ``spinless representation'' trick used to extract spin blocks from a doubled-orbital tensor.
%     \item \texttt{pyscf/fci/rdm.py}: density-matrix kernels and post-processing utilities, notably \texttt{make\_dm1234} and \texttt{reorder\_dm1234}, which compute and reorder 1--4 RDMs in canonical convention.
% \end{itemize}
% Crucially, \texttt{direct\_spin1.py} contains commented algebraic consistency asserts for 2- and 3-RDMs that specify how spin-separated tensors sum (with transpositions) into the spin-traced tensor; these served as ``indexing breadcrumbs'' to generalize the reconstruction logic to the 4-RDM case.

\medskip
{\normalsize\textbf{Implementation Approach}}\\
The solution mirrors the established design patterns presented by the minimal API spec, as well as the pattern present in
\texttt{pyscf/fci/direct\_spin1.py} for lower-order RDMs:
\medskip

\begin{enumerate}[nosep,leftmargin=*]
    \item \textbf{Spin-traced 1--4 RDMs.}  
    The implementation delegates all heavy computation to the existing kernel
    \texttt{rdm.make\_dm1234}, followed by an optional call to
    \texttt{rdm.reorder\_dm1234} to enforce PySCF’s canonical operator ordering.

    \item \textbf{Spin-separated 1--4 RDMs.}  
    As in \texttt{make\_rdm123s}, the CI vector is first converted into a
    spinless representation over doubled orbitals. The spin-traced RDMs are then
    computed in this enlarged space and sliced into $\alpha/\beta$ blocks using
    spatial index ranges. The slicing logic generalizes the existing 2- and
    3-RDM implementations to the full set of 3- and 4-RDM spin blocks.
\end{enumerate}

% \begin{tcolorbox}[
%     colback=green!5,
%     colframe=green!40!black,
%     left=2pt,right=2pt,top=2pt,bottom=2pt,
%     boxrule=0.4pt,
%     width=0.92\linewidth,
%     enhanced jigsaw,
%     breakable,
% ]
% \scriptsize
% \begin{Verbatim}[breaklines=true, breakanywhere=true]

\vspace{2pt}
\noindent
\textit{Representative excerpt (truncated):}
\begin{tcolorbox}[
    colback=green!5,
    colframe=green!40!black,
    left=2pt,right=2pt,top=2pt,bottom=2pt,
    boxrule=0.4pt,
    width=0.92\linewidth,
    enhanced jigsaw,
    breakable,
]
\scriptsize
\begin{Verbatim}[breaklines=true, breakanywhere=true, commandchars=\\\{\}]
# pyscf/fci/direct_spin1.py  (excerpt; code truncated)

def make_rdm1234(fcivec, norb, nelec, link_index=None, reorder=True):
    dm1, dm2, dm3, dm4 = rdm.make_dm1234(...)
    if reorder:
        dm1, dm2, dm3, dm4 = rdm.reorder_dm1234(...)
    return dm1, dm2, dm3, dm4

def make_rdm1234s(fcivec, norb, nelec, link_index=None, reorder=True):
    ci_spinless = civec_spinless_repr(...)
    rdm1, rdm2, rdm3, rdm4 = make_rdm1234(...)
    # slice \( \alpha/\beta \) blocks for 1–4 RDMs
    # ...
    return (rdm1a, rdm1b), (rdm2aa, rdm2ab, rdm2bb), ...
\end{Verbatim}
\end{tcolorbox}

\noindent

\medskip
{\normalsize\textbf{Significance}}\\
This case illustrates how high-leverage scientific extensions can be driven by surprisingly small surface specifications when the codebase's naming conventions and helper utilities are coherent. By aligning with PySCF's established RDM conventions and reusing \texttt{rdm.make\_dm1234}+\texttt{rdm.reorder\_dm1234}, the implementation cleanly extends the FCI solver to provide spin-resolved access to 3- and 4-body correlators. In practice, these APIs ensure that downstream methods can reliably reconstruct spin-traced tensors.

\end{casebox}

\begin{casebox}[top=2pt,bottom=2pt,left=4pt,right=4pt,breakable]{
\vspace{-0.6\baselineskip}
\subsection{False Positive: Chirality Constraints in RDKit Embedding}}
\scriptsize   % 比 \small 再小一档，整体更紧凑

\textbf{Pull Request Description.}

\medskip
The RDKit conformation generator \texttt{EmbedMolecule()} should fail when a molecule
contains a geometrically impossible tetrahedral center. In PR~\#7564, certain molecules
return \(-1\), indicating failure to embed. The maintainer-provided fix introduces a new
\texttt{d\_structureFlags} field and conditionally relaxes the tetrahedral volume threshold
\textit{only} for chiral centers located in fused small rings (high angular strain).

\medskip

{\scriptsize
\textbf{Model-Generated Patch (Key Diff Excerpts).}

% --- Incorrect Model Patch (False Positive) ---
\begin{tcolorbox}[
    colback=green!5,
    colframe=green!40!black,
    left=2pt,right=2pt,top=2pt,bottom=2pt,   % padding 更小
    boxrule=0.4pt,                           % 细一点的边框
    width=0.92\linewidth                     % ★ 同样：0.92\linewidth
]
\scriptsize
\begin{verbatim}
+ constexpr double MIN_TETRAHEDRAL_CHIRAL_VOL = 0.20;
+ constexpr double TETRAHEDRAL_CENTERINVOLUME_TOL = 0.15;

+ // overly strict linearity tolerance
+ double linearTol = 1e-5;

+ // skip hydrogens (not part of official logic)
+ if (nbr->getAtomicNum() == 1) {
+     continue;
+ }
\end{verbatim}
\end{tcolorbox}
}

\medskip
\textbf{What did the model do?}
The model produced a patch that globally relaxes two critical constants:
\begin{itemize}
    \item \texttt{MIN\_TETRAHEDRAL\_CHIRAL\_VOL}: \(0.50 \rightarrow 0.20\)
    \item \texttt{TETRAHEDRAL\_CENTERINVOLUME\_TOL}: \(0.30 \rightarrow 0.15\)
\end{itemize}

It also adds minor changes unrelated to the scientific logic (such as hydrogens being
skipped in linearity checks), but \textit{does not} modify the \texttt{ChiralSet} data
structure and \textbf{does not implement structural-context--dependent logic}. 

\medskip
\textbf{Why is this a false positive?}
Under the benchmark test suite, this globally relaxed version of the embedding algorithm
passes because some previously failing SMILES strings now embed successfully. However, this constitutes a \textit{scientific regression}. Molecules
with geometrically impossible tetrahedral centers, such as:
\begin{verbatim}
O=C(O)[C@H](F)=C(F)F
CC([C@H](Cl)=C(Cl)Cl)F
C1CC1[C@H](F)Cl
\end{verbatim}
should \textit{not} be embeddable. The model's patch embeds them by lowering global
thresholds rather than addressing the actual geometric constraints. 

\medskip
\textbf{Scientific intent vs. superficial test passing.}
The test suite introduces bulk test cases for valid small-ring embeddings but does not
include dedicated “impossible geometry” regressions. The model’s incorrect
patch passes all tests even though it fundamentally violates the intended physical
meaning of the embedding algorithm. This is a canonical example of a \textit{false
positive}: a patch that satisfies the unit tests while failing the scientific semantics.

\tcblower
\scriptsize
\textbf{Case Study Description.}
A globally relaxed chirality threshold makes impossible molecules embeddable, letting the model’s patch pass tests while violating physical constraints. This shows how scientific software can yield false positives without domain-aware checks.
\label{case:chiral-fp}
\end{casebox}

\begin{casebox}[top=2pt,bottom=2pt,left=4pt,right=4pt,breakable]{
\vspace{-0.6\baselineskip}
\subsection{False Negative: Overly Strict DIIS Iteration Cap in PySCF ROHF}}

\scriptsize

\textbf{Pull Request Description.}

\medskip

This pull request addresses a DIIS instability for a symmetry-adapted ROHF calculation on atomic oxygen:
with DIIS enabled, the triplet O state (\texttt{spin=2}, \texttt{basis='cc-pVDZ'}, \texttt{symmetry=True}) fails to converge, whereas disabling DIIS yields a stable solution.
The maintainer fix modifies DIIS error-vector construction and symmetry handling (e.g., in \texttt{diis.py}, \texttt{hf\_symm.py}, \texttt{uhf\_symm.py}, \texttt{ghf\_symm.py}) so that symmetry-forbidden components are masked out and symmetry-adapted SCF paths consistently require \texttt{mol.symmetry} to be enabled.
The corresponding test \texttt{test\_diis\_for\_symmetry\_adapted\_scf} asserts that ROHF with DIIS now converges for triplet O \textit{within} a fixed maximum of 9 SCF cycles and reaches a reference energy of \(-74.7874921601008\)~Ha.

\medskip

{\scriptsize
\textbf{Model-Generated Patch (Key Diff Excerpt, from \texttt{rohf.get\_fock}).}

\begin{tcolorbox}[
    colback=green!5,
    colframe=green!40!black,
    left=2pt,right=2pt,top=2pt,bottom=2pt,
    boxrule=0.4pt,
    width=0.92\linewidth
]
\scriptsize
\begin{verbatim}
-   if diis and cycle >= diis_start_cycle:
-       f = diis.update(s1e, dm_tot, f, mf, h1e, vhf)
+   if diis and cycle >= diis_start_cycle:
+       // perform DIIS on spin-resolved Fock matrices, not on the
+       // Roothaan effective Fock
+       auto fdiis = mf._rohf_fock_diis;
+       if (!fdiis) {
+           fdiis = std::make_shared<EDIIS>(mf, mf.diis_file);
+           fdiis->space = mf.diis_space;
+           mf._rohf_fock_diis = fdiis;
+       }
+       // mix (Fa, Fb) directly and then rebuild the Roothaan Fock
+       fdiis->Corth = nullptr;
+       auto fab = fdiis->update(s1e, dm, make_array(focka, fockb), mf, h1e, vhf);
+       focka = fab[0];
+       fockb = fab[1];
+       f = get_roothaan_fock(std::make_pair(focka, fockb), dm, s1e);
\end{verbatim}
\end{tcolorbox}
}

\medskip

\textbf{What did the model do?}

The model left the maintainer’s symmetry-handling code unchanged but altered how DIIS is applied in ROHF:
instead of mixing the single “Roothaan” effective Fock matrix, it performs EDIIS mixing directly on the spin-resolved Fock matrices \((F_\alpha, F_\beta)\) and reconstructs the effective Fock afterward.
In the reproduced failing case (triplet O, \texttt{cc-pVDZ}, \texttt{symmetry=True}, \texttt{init\_guess='1e'}, constrained \texttt{irrep\_nelec}), this modified DIIS scheme converges reliably to the same total energy as the maintainer patch, matching the reference energy to at least 9 decimal places and yielding \texttt{mf.converged == True}.
However, it typically requires around 16 SCF iterations rather than 9.

\medskip

\textbf{Why is this a false negative?}

Under the benchmark’s test patch, \texttt{test\_diis\_for\_symmetry\_adapted\_scf} sets \texttt{mf.max\_cycle = 9} and asserts both convergence and agreement with the reference energy.
The model’s patch \textit{does} restore DIIS convergence for the problematic ROHF calculation and reaches the correct physical solution, but only after more than 9 iterations.
Consequently, the test fails not because the scientific behavior is incorrect, but because the test encodes an arbitrary performance constraint (convergence within 9 cycles) that was never part of the original bug report.
This is therefore a \textit{false negative}: a valid scientific fix is rejected solely due to an overly strict iteration cap in the test, rather than any discrepancy in the final wavefunction or energy.

\medskip

\textbf{Scientific intent vs. test design.}

The scientific requirement the issue is that ROHF with DIIS should converge for the specified triplet O system and reproduce the correct energy, just as the DIIS-disabled calculation does.
The model’s patch satisfies these physical constraints but deviates from the maintainer’s chosen implementation path and convergence speed.
Because the test suite over-specifies iteration constraints, it effectively enforces one particular implementation style rather than the underlying quantum-chemical behavior.

\tcblower

\scriptsize

\textbf{Case Study Description/Solution}

A ROHF-specific DIIS redesign in PySCF restores convergence for a problematic triplet oxygen calculation and yields the correct total energy, but fails a benchmark test that \textit{requires convergence within 9 iterations}. In this case fixing the false negative is quite simple, and requires only that we remove this the 9 iteration requirement.

\label{case:diis-fn}

\end{casebox}

\begin{casebox}[top=2pt,bottom=2pt,left=4pt,right=4pt,breakable]{
\vspace{-0.6\baselineskip}
\subsection{Implement truncated octahedral and rhombic dodecahedral solvation}}\refstepcounter{case}\label{case:openmm-solvent-geometry} \scriptsize

{\normalsize\textbf{The Task}}

The agent was asked to extend OpenMM’s \texttt{Modeller.addSolvent()} method to support non-cubic periodic boxes, specifically truncated octahedra and rhombic dodecahedra. The implementation needed to correctly calculate periodic box vectors based on a user-defined padding distance $p$, ensuring that the minimum distance between any point of the solute and its periodic images in all directions is at least $p$.

\medskip
{\normalsize\textbf{Physical Context}}\\
Molecular dynamics simulations use periodic boundary conditions (PBC) to simulate bulk properties. While cubic boxes are standard, compact shapes like the truncated octahedron (the Wigner–Seitz cell of a BCC lattice) and the rhombic dodecahedron (the Voronoi cell of an FCC lattice) are more computationally efficient. They require ~23\% and ~29\% less volume respectively than a cube for the same minimum image distance, reducing the number of solvent molecules needed.

For non-orthogonal boxes, a naive axis-aligned bounding box (AABB) approach is insufficient. If a solute is elongated along a non-axial direction, the distance to a tilted face may be smaller than the distance to an axial boundary. To guarantee a minimum padding $p$, the box must be sized relative to the solute's bounding sphere of radius $R$. The periodic box width $L$ must then satisfy $L = 2R + p$ (or $2R + 2p$ depending on the convention) to ensure no part of the solute interacts with its own image within the cutoff distance.

\medskip
{\normalsize\textbf{Domain Knowledge}}\\
Key concepts required for this implementation include:

\begin{enumerate}[nosep,leftmargin=*]
    \item \textit{Bounding Sphere vs. AABB}: Sizing a box based on $max(x, y, z)$ (AABB) only guarantees padding along the axes. For triclinic boxes, the minimal distance to a face often occurs off-axis, necessitating a sphere-based radius $R = \max \|\vec{r}_i - \vec{c}\|$.
    \item \textit{Reduced Triclinic Vectors}: OpenMM requires box vectors in a "reduced" form where vector $\vec{a}$ is along $x$, $\vec{b}$ is in the $xy$-plane, and components are shifted to satisfy $b_x \le a_x/2$, etc.
    \item \textit{Shape Geometry}: A truncated octahedron has angles $\alpha=\beta=\gamma=\arccos(-1/3) \approx 109.47^\circ$. A rhombic dodecahedron typically uses $60^\circ$ or $120^\circ$ angles.
    \item \textit{Volume/Density Heuristics}: When adding a specific number of molecules, the volume of the shape ($V_{TO} = L^3/2$; $V_{RD} = L^3/\sqrt{2}$) must be accounted for to estimate the correct padding.
\end{enumerate}

[Image comparing bounding sphere and axis-aligned bounding box for a non-spherical solute]

\medskip
{\normalsize\textbf{Reference Answer}}\\
The correct implementation calculates the bounding sphere radius and applies shape-specific coefficients to generate reduced vectors:

\medskip

{\scriptsize
\textbf{Model-Generated Patch (Key Diff Excerpts).}
\begin{tcolorbox}[
    colback=green!5,
    colframe=green!40!black,
    left=2pt,right=2pt,top=2pt,bottom=2pt,
    boxrule=0.4pt,
    width=0.92\linewidth,
    enhanced jigsaw,
    breakable,
]
\scriptsize
\begin{Verbatim}[breaklines=true, breakanywhere=true]
# 1. Compute bounding sphere
center = 0.5 * (minRange + maxRange)
radius = max(unit.norm(center - pos) for pos in positions)
width = 2 * radius + padding

# 2. Define reduced vectors based on width
if boxShape == 'cube':
    vectors = (Vec3(width, 0, 0), Vec3(0, width, 0), Vec3(0, 0, width))
elif boxShape == 'dodecahedron':
    vectors = (Vec3(width, 0, 0), Vec3(0, width, 0), Vec3(0.5, 0.5, 0.5*sqrt(2))*width)
elif boxShape == 'octahedron':
    vectors = (Vec3(width, 0, 0), Vec3(1/3, 2*sqrt(2)/3, 0)*width, Vec3(-1/3, sqrt(2)/3, sqrt(6)/3)*width)
\end{Verbatim}
\end{tcolorbox}
}
\medskip {\normalsize\textbf{Failure patterns}}\ The agent correctly identified the necessary API changes but failed the geometric logic. It used the axis-aligned extent (\texttt{maxSize}) and then attempted to use a generic \texttt{computePeriodicBoxVectors} function with angles.

\scriptsize
\textbf{Model-Generated Patch (Key Diff Excerpts).}
\begin{tcolorbox}[
    colback=green!5,
    colframe=green!40!black,
    left=2pt,right=2pt,top=2pt,bottom=2pt,
    boxrule=0.4pt,
    width=0.92\linewidth,
    enhanced jigsaw,
    breakable,
]
\scriptsize
\begin{Verbatim}[breaklines=true, breakanywhere=true]
# Agent's incorrect logic

maxSize = max(max(pos[i]) - min(pos[i]) for i in range(3)) width = maxSize + 2*padding 
if boxShape == 'octahedron': 
    vectors = computePeriodicBoxVectors(width, width, width, 109.47, 109.47, 109.47) 
    box = Vec3(vectors[0][0], vectors[1][1], vectors[2][2]) 

# Error: treats box as rectangular 
\end{Verbatim}
\end{tcolorbox}

This fails for two reasons: (1) \texttt{maxSize} (AABB) underestimates the required box size for non-spherical solutes in tilted boxes. (2) The agent treated the resulting vectors as a rectangular box for molecule inclusion tests (\texttt{any(atomPos[i] > box[i])}), which is invalid for non-orthogonal vectors and would lead to incorrect solvent placement near the skewed boundaries.

\medskip { {\normalsize\textbf{Significance}}\ This is a "geometric algorithmic miscalculation." The agent's code is syntactically correct and uses the right library functions, but it violates the underlying physical requirement of the task (the padding guarantee). In scientific computing, such errors are dangerous because the simulation will run without crashing, but the physics will be wrong. In this case, it could causing artificial and physically incorrect interactions between the solute and its periodic images. This highlights that LLMs must understand the \textit{geometric constraints} of physical systems.} \end{casebox}

\begin{casebox}[top=2pt,bottom=2pt,left=4pt,right=4pt,breakable]{
\vspace{-0.6\baselineskip}
\subsection{Fix Bloch-phase convention in k-point symmetry-adapted periodic CCSD}}\refstepcounter{case}\label{case:pyscf-bloch-phase-sign} \scriptsize

{\normalsize\textbf{The Task}}\\
Coupled-cluster singles and doubles (CCSD) is an ab initio many-body electronic structure method widely used for high-accuracy correlation energy calculations. The agent was asked to diagnose and fix a non-convergence bug in PySCF’s periodic CCSD implementation that exploits k-point space-group symmetry (\texttt{pyscf.pbc.cc.kccsd\_rhf\_ksymm.KsymAdaptedRCCSD}). A user reported that a diamond primitive cell with a $3\times 3\times 3$ k-point mesh fails to converge when k-point symmetry is enabled, while the non-symmetry-adapted \texttt{KRCCSD} converges. The goal was to identify the scientific root cause in the symmetry machinery that maps orbitals and integrals between the irreducible Brillouin zone (IBZ) and the full Brillouin zone (BZ), and to implement the minimal corrective change so that the CCSD calculation converges.

\medskip
{\normalsize\textbf{Physical Context}}\\
In periodic electronic structure theory, crystalline orbitals are Bloch functions $\psi_{n,\mathbf{k}}(\mathbf{r})$ obeying Bloch’s theorem
\[
\psi_{n,\mathbf{k}}(\mathbf{r}+\mathbf{R}) = e^{i\,\mathbf{k}\cdot\mathbf{R}}\,\psi_{n,\mathbf{k}}(\mathbf{r})
\quad \text{for any lattice vector } \mathbf{R}.
\]
Space-group operations $g\equiv (\mathcal{R}\mid \boldsymbol{\tau})$ act as $\mathbf{r}\mapsto \mathcal{R}\mathbf{r}+\boldsymbol{\tau}$ and induce $\mathbf{k}\mapsto \mathcal{R}\mathbf{k}$. When generating symmetry-related quantities from the IBZ to the full BZ, one must include the \textit{correct} Bloch phase associated with any net lattice translation $\mathbf{L}$ incurred when mapping atomic centers between symmetry-equivalent positions. If the sign convention in the phase $e^{\pm i\,\mathbf{k}\cdot\mathbf{L}}$ is wrong, the transformed orbitals/integrals acquire inconsistent complex phases across the star of k-points, breaking Hermiticity and momentum-conservation relationships that CCSD relies on.

A key diagnostic is the MP2 correlation energy used to initialize CCSD:
\[
E^{(2)} = \sum_{ijab,\mathbf{k}} 
\frac{|\langle ij\Vert ab\rangle|^2}{\epsilon_i+\epsilon_j-\epsilon_a-\epsilon_b}.
\]
For an insulator, $\epsilon_i,\epsilon_j$ (occupied) lie below $\epsilon_a,\epsilon_b$ (virtual), making denominators negative and yielding a \textit{negative} $E^{(2)}$. With an incorrect symmetry phase, the ERIs can become pathologically inconsistent, producing unphysical MP2 energies (including positive values and large imaginary parts), and CCSD iterations can fail to converge.

\medskip
{\normalsize\textbf{Domain Knowledge}}\\
Key concepts required for this diagnosis include:
\begin{enumerate}[nosep,leftmargin=*]
    \item \textit{Bloch phase under translations}: the canonical convention is $e^{+i\,\mathbf{k}\cdot\mathbf{L}}$ in $\psi_{\mathbf{k}}(\mathbf{r}+\mathbf{L})$; using the complex conjugate globally misaligns phases across k-point stars.
    \item \textit{Space-group actions and IBZ$\leftrightarrow$BZ reconstruction}: mapping orbitals/integrals between symmetry-related k-points requires consistent rotation of $\mathbf{k}$ plus translation-induced phases tied to how atoms are permuted and shifted by lattice vectors.
    \item \textit{Hermiticity and momentum conservation in k-space ERIs}: correct symmetry reconstruction preserves algebraic constraints; phase errors corrupt complex ERIs and can induce large imaginary components in energies/intermediates.
    \item \textit{Energy-sign sanity checks}: a positive MP2 correlation energy in a gapped system is a red flag for inconsistent integrals/phase conventions rather than mere orbital ordering.
\end{enumerate}

\medskip
{\normalsize\textbf{Reference Answer}}\\
The root cause was upstream of CCSD: the Bloch phase sign used during symmetry transformations between the IBZ and the full BZ was incorrect. In PySCF this occurs in \texttt{pyscf/pbc/symm/symmetry.py} (routine \texttt{\_get\_phase}), which computes phase factors associated with mapping atomic centers under a space-group operation, including a lattice shift \texttt{Lshift}. The original code used the complex-conjugated phase,
\[
e^{-i\,2\pi\,\mathbf{k}\cdot\mathbf{L}},
\]
where $\mathbf{k}$ is in scaled reciprocal coordinates. The gold patch flips the sign to match Bloch’s theorem:
\[
e^{+i\,2\pi\,\mathbf{k}\cdot\mathbf{L}}.
\]
This one-line fix restores consistent phases for orbitals/integrals generated from the IBZ, yielding physically reasonable (negative) MP2 initialization energy and eliminating the spurious imaginary components that destabilized CCSD convergence.

\begin{tcolorbox}[
    colback=green!5,
    colframe=green!40!black,
    left=2pt,right=2pt,top=2pt,bottom=2pt,
    boxrule=0.4pt,
    width=0.92\linewidth,
    enhanced jigsaw,
    breakable,
]
\scriptsize
\begin{Verbatim}[breaklines=true, breakanywhere=true]
# pyscf/pbc/symm/symmetry.py  (in _get_phase)

# remove numerical noise, important for symmetry adaptation
Lshift = Lshift.round()
if not ignore_phase:
-    phase[iatm] = np.exp(-1j * np.dot(kpt_scaled, Lshift) * 2.0 * np.pi)
+    phase[iatm] = np.exp( 1j * np.dot(kpt_scaled, Lshift) * 2.0 * np.pi)
\end{Verbatim}
\end{tcolorbox}

\medskip
{\normalsize\textbf{Failure patterns}}\\
The agent misdiagnosed the symptom (non-convergence and unphysical MP2 energy) as an orbital-ordering issue and patched the CCSD ERI builder instead of the symmetry phase convention. Concretely, it modified \texttt{pyscf/pbc/cc/kccsd\_rhf\_ksymm.py} to compute k-point Fock matrices, take their diagonals as \texttt{mo\_energy}, and then sort \texttt{mo\_energy} and the corresponding \texttt{mo\_coeff} columns when not monotone.

\begin{tcolorbox}[
    colback=green!5,
    colframe=green!40!black,
    left=2pt,right=2pt,top=2pt,bottom=2pt,
    boxrule=0.4pt,
    width=0.92\linewidth,
    enhanced jigsaw,
    breakable,
]
\scriptsize
\begin{Verbatim}[breaklines=true, breakanywhere=true]
# Agent's incorrect patch (excerpt)
mo_energy = [fock[k].diagonal().real for k in range(len(fock))]

# Ensure mo_energy and mo_coeff are sorted
for k in range(len(mo_energy)):
    if not np.all(mo_energy[k][:-1] <= mo_energy[k][1:]):
        logger.warn(cc, 'MO energies not sorted at k-point %d. Sorting...', k)
        idx = np.argsort(mo_energy[k])
        mo_energy[k] = mo_energy[k][idx]
        mo_coeff[k] = mo_coeff[k][:, idx]
        fock[k] = fock[k][idx][:, idx]
self.fock = fock
self.mo_coeff = mo_coeff
\end{Verbatim}
\end{tcolorbox}

This does not (and cannot) repair a \textit{global} phase inconsistency across symmetry-related k-points: sorting only reorders eigenvectors locally at each k-point, while the true error corrupts the inter-k-point phase relationships that define the symmetry-adapted reconstruction of orbitals and ERIs. In degenerate manifolds, such sorting may further introduce arbitrary gauge choices, potentially masking the real defect and destabilizing reproducibility.

\medskip
{\normalsize\textbf{Significance}}\\
This case highlights a characteristic failure mode in scientific software repair: the code can be modified in a superficially reasonable way (``orbital energies aren’t sorted''), yet still miss a subtle but foundational physics convention. Here, a single sign in the Bloch phase $e^{\pm i\mathbf{k}\cdot\mathbf{L}}$ determines whether IBZ$\rightarrow$BZ transformations preserve the complex phase structure required for Hermiticity and momentum conservation. With the wrong sign, downstream correlated methods exhibit dramatic, qualitative pathologies (positive MP2 energies in an insulator, large imaginary parts, and CCSD non-convergence). The episode illustrates that effective debugging in computational chemistry often hinges on recognizing invariants and conventions that are only weakly encoded in unit tests or APIs; without that domain grounding, even clean-looking patches can be scientifically inert or actively harmful.

\end{casebox}

\begin{casebox}[top=2pt,bottom=2pt,left=4pt,right=4pt,breakable]{
\vspace{-0.6\baselineskip}
\subsection{RHF smearing must conserve electron number for odd $N$}}\refstepcounter{case}\label{case:pyscf-rhf-smearing-odd-nelec} \scriptsize

{\normalsize\textbf{The Task}}

The agent was asked to fix PySCF’s smearing implementation for restricted Hartree--Fock (RHF) so that the total electron number is conserved for systems with an odd number of electrons. The bug appears when \texttt{pyscf.scf.addons.smearing} is applied to an RHF object with odd \texttt{mol.nelectron}: the resulting occupations \texttt{mo\_occ} sum to an \textit{even} integer (e.g., 16) instead of the correct odd electron count (e.g., 15). The relevant code lives in \texttt{pyscf/scf/addons.py}, specifically \texttt{\_SmearingSCF.get\_occ} (RHF branch) and the helper \texttt{\_get\_fermi} used to seed the chemical potential $\mu$.

\medskip
{\normalsize\textbf{Physical Context}}\\
In Hartree--Fock theory, electrons occupy molecular orbitals (MOs) determined by a self-consistent Fock operator. In \textit{restricted} HF, $\alpha$ and $\beta$ spins share the same spatial orbitals, so each spatial MO carries a two-fold spin degeneracy and can host up to two electrons. At zero temperature, RHF occupations are typically $0$ or $2$, but smearing introduces fractional occupations to stabilize SCF convergence by approximating a finite-temperature ensemble.

For Fermi--Dirac smearing, the (spin-resolved) occupation of an orbital with energy $\varepsilon_i$ is
\[
f_i(\mu,\sigma) = \frac{1}{\exp\!\big((\varepsilon_i-\mu)/\sigma\big)+1},
\]
(and PySCF also supports a Gaussian-like form, e.g.\ $\tfrac{1}{2}\,\mathrm{erfc}((\varepsilon_i-\mu)/\sigma)$). The chemical potential $\mu$ must be chosen to enforce particle-number conservation. For RHF, the constraint is
\[
2\sum_i f_i(\mu,\sigma) = N,
\]
so the \textit{spatial-orbital} occupation sum must equal $N/2$, which is fractional when $N$ is odd. If one instead rounds the spatial occupation count to an integer and then doubles, the total becomes even, violating particle-number conservation.

\medskip
{\normalsize\textbf{Domain Knowledge}}\\
Key concepts required for this fix include:
\begin{enumerate}[nosep,leftmargin=*]
    \item \textit{RHF degeneracy and electron counting}: enforce $\sum_i n_i = N$ with $n_i \in [0,2]$ per spatial orbital, i.e.\ solve the smearing constraint in the spatial basis with target $N/2$.
    \item \textit{Chemical potential as a constraint variable}: $\mu$ is not a free parameter; it must be adjusted so the occupation function satisfies the electron-number constraint.
    \item \textit{Fractional $N/2$ for odd-electron RHF}: odd $N$ implies a half-filled spatial level \textit{on average}, which requires allowing non-integer $N/2$ in the solver.
    \item \textit{Robust initialization of $\mu$}: seeding $\mu$ near the HOMO/LUMO region requires turning a fractional target (e.g.\ $7.5$) into a valid index, e.g.\ via $\lceil N/2 \rceil$.
    \item \textit{Consistent entropy scaling}: when occupations are computed per spin channel and then doubled for RHF, the entropy (free-energy term) must be doubled as well.
\end{enumerate}

\medskip
{\normalsize\textbf{Reference Answer}}\\
The correct fix enforces the RHF electron-number constraint in the spatial basis and correctly handles fractional $N/2$ when seeding $\mu$.

\begin{tcolorbox}[
    colback=green!5,
    colframe=green!40!black,
    left=2pt,right=2pt,top=2pt,bottom=2pt,
    boxrule=0.4pt,
    width=0.92\linewidth,
    enhanced jigsaw,
    breakable,
]
\scriptsize
\begin{Verbatim}[breaklines=true, breakanywhere=true]
# In pyscf/scf/addons.py

def _get_fermi(mo_energy, nocc):
    mo_e_sorted = numpy.sort(mo_energy)
    # nocc may be fractional (e.g., 7.5), so use ceil to pick a valid index
    return mo_e_sorted[numpy.ceil(nocc).astype(int) - 1]

...

# In _SmearingSCF.get_occ (RHF branch)
if is_rhf:
    nocc = nelectron / 2  # fractional target allowed for odd N

mu, mo_occs = _smearing_optimize(f_occ, mo_es, nocc, sigma)
self.entropy = self._get_entropy(mo_es, mo_occs, mu)

if is_rhf:
    mo_occs *= 2
    self.entropy *= 2
\end{Verbatim}
\end{tcolorbox}

With this change, a 15-electron system targets $\sum_i f_i = 7.5$ in the spatial basis and, after doubling, yields \texttt{mo\_occ.sum() = 15} exactly.

\medskip
{\normalsize\textbf{Failure patterns}}\\
The agent’s patch preserved an integer-rounded spatial occupation target in RHF:
\begin{tcolorbox}[
    colback=green!5,
    colframe=green!40!black,
    left=2pt,right=2pt,top=2pt,bottom=2pt,
    boxrule=0.4pt,
    width=0.92\linewidth,
    enhanced jigsaw,
    breakable,
]
\scriptsize
\begin{Verbatim}[breaklines=true, breakanywhere=true]
def _get_fermi(mo_energy, nocc):
    mo_e_sorted = numpy.sort(mo_energy)
    return mo_e_sorted[nocc-1]  # assumes integer nocc

...

if is_rhf:
    nocc = (nelectron + 1) // 2  # integer rounding

mu, mo_occs = _smearing_optimize(f_occ, mo_es, nocc, sigma)

if is_rhf:
    mo_occs *= 2
\end{Verbatim}
\end{tcolorbox}

For odd $N$, this guarantees an even result: e.g.\ $N=15 \Rightarrow (N+1)//2=8$, so the solver enforces $\sum_i f_i = 8$ and then doubles to 16. The code is numerically stable, compiles, and can converge, but it violates the conservation law the method is supposed to enforce.

\medskip
{\normalsize\textbf{Significance}}\\
This case is a prototypical \textit{conservation-law violation} triggered by an ``innocent'' engineering choice (integer rounding) that is valid for closed-shell even-$N$ RHF but becomes unphysical under smearing for odd-electron systems. The failure is subtle: SCF still converges and most regression tests (often dominated by even-electron molecules) can pass, yet the algorithm breaks a fundamental invariant ($N$ conservation). The fix requires explicitly reasoning about RHF spin degeneracy and recognizing that $N/2$ is \textit{allowed to be fractional} in the smeared ensemble, illustrating how domain constraints govern scientific software validity.

\end{casebox}

\begin{casebox}[top=2pt,bottom=2pt,left=4pt,right=4pt,breakable]{
\vspace{-0.6\baselineskip}
\subsection{Order-independent tautomer hashing in RDKit MolHash}}\refstepcounter{case}\label{case:rdkit-tautomerhashv2-order} \scriptsize

{\normalsize\textbf{The Task}}\\
The agent was asked to diagnose and fix an order-dependence bug in RDKit’s tautomer hash function \texttt{HetAtomTautomerv2} (implemented in \texttt{Code/GraphMol/MolHash/hashfunctions.cpp}). The bug manifested as different hash outputs for the \textit{same} molecule when constructed from a V3000 molblock versus from canonical SMILES, indicating that traversal/bond iteration order (or atom numbering) was leaking into the hash. The requirement was to make \texttt{MolHash::TautomerHashv2} deterministic and canonical (independent of bond order and atom indices) while preserving scientific correctness about which atoms/bonds belong to tautomeric systems and maintaining existing semantics and tests.

\medskip
{\normalsize\textbf{Physical Context}}\\
Tautomerism is a chemically constrained rearrangement involving proton relocation and adjacent $\pi$-bond shifts (e.g., keto--enol). A ``tautomeric hash'' aims to produce a canonical identifier for all members of a tautomer set: equivalent tautomers should map to the same hash, while chemically distinct molecules should not be spuriously merged. Computationally, the algorithm identifies a subgraph $G=(V,E)$ of \textit{eligible} bonds and selects a subset $S\subseteq E$ representing tautomeric/conjugated systems under domain constraints (conjugation, heteroatom behavior, and functional group exclusions). It then normalizes that system (e.g., aromaticization of bonds in $S$, charge/H bookkeeping), and emits a canonical SMILES-like representation with additional invariants (e.g., H-count / net charge suffixes). Determinism is not a pure ``graph ordering'' property: it must arise from chemically-defined inclusion/exclusion rules so that $S$ is stable across alternate atom numbering and input encodings.

\medskip
{\normalsize\textbf{Domain Knowledge}}\\
Key concepts required for a correct fix include:
\begin{enumerate}[nosep,leftmargin=*]
    \item \textit{Conjugated system identification}: Tautomeric motion occurs along conjugated pathways, so inclusion decisions must depend on conjugation predicates (e.g., conjugated bond neighborhoods), not on incidental traversal order.
    \item \textit{Functional group exclusions via SMARTS flags}: Many motifs (amides, nitro, phosphates, sulfates, guanidines, isocyanates, etc.) must \textit{not} be treated as tautomerizing; these are excluded by SMARTS-driven bond flags used during traversal.
    \item \textit{Overreach prevention near heteroatoms}: In patterns like enamine-like \texttt{C--N--C=C}, naive expansion can incorrectly absorb the first adjacent single bond near a heteroatom. A targeted overreach heuristic based on local conjugation density prevents chemically invalid system growth.
    \item \textit{Flag-aware eligibility}: Atom participation tests must respect propagated atom/bond flags (e.g., using \texttt{isCandidateAtom(oatom, atomFlags)} rather than an overload that ignores flags).
    \item \textit{Determinism via chemistry, not sorting}: Robust canonicalization requires stabilizing \textit{which} bonds are selected (chemistry), not merely stabilizing \textit{how} bonds are iterated (data structure order).
\end{enumerate}

\medskip
{\normalsize\textbf{Reference Answer}}\\
The reference solution stabilizes \texttt{TautomerHashv2} by \textit{augmenting chemical heuristics} rather than applying generic ordering tricks. Concretely, it:
(i) extends \texttt{getBondFlags} with additional SMARTS patterns to flag/exclude non-tautomeric functional groups; 
(ii) introduces a per-atom \texttt{getNumConjugatedNeighbors()} statistic to quantify local conjugation density;
(iii) implements \texttt{checkForOverreach()} to block specific heteroatom-adjacent expansions; and
(iv) integrates these checks into neighbor exploration, deduplicating queued neighbor bonds, tracking conjugated atoms, and using flag-aware candidate checks.

\begin{tcolorbox}[
    colback=green!5,
    colframe=green!40!black,
    left=2pt,right=2pt,top=2pt,bottom=2pt,
    boxrule=0.4pt,
    width=0.92\linewidth,
    enhanced jigsaw,
    breakable,
]
\scriptsize
\begin{Verbatim}[breaklines=true, breakanywhere=true]
# Sketch of gold approach (C++ in hashfunctions.cpp)
startBonds[b] = isPossibleStartingBond(b, atomFlags, bondFlags)

for atm in mol->atoms():
    numConjNbrs[atm] = getNumConjugatedNeighbors(atm, startBonds)

BFS/DFS over eligible bonds:
  if (checkForOverreach(atm, oatom, bptr, nbrBond,
                        startBonds, numConjNbrs)) continue;
  if (skipNeighborBond(..., atomFlags, bondFlags)) continue;
  if (nbrBond not already in queue) queue.push(nbrBond);
  mark conjSystem, conjAtoms; use isCandidateAtom(oatom, atomFlags)
\end{Verbatim}
\end{tcolorbox}

These changes make the selected tautomeric system $S$ chemically stable across representations, yielding deterministic hashes without breaking chemically meaningful distinctions.

\medskip
{\normalsize\textbf{Failure patterns}}\\
The agent misdiagnosed a chemistry-driven selection problem as a generic iteration-order problem, and attempted to enforce determinism by sorting and cardinality-based set selection. It (a) sorted bonds by atom indices, (b) re-initialized \texttt{bondsConsidered} per traversal seed, and (c) chose systems using either a union-of-all strategy or a ``pick smallest disjoint systems'' heuristic. None of these encode the required chemical constraints (exclusions, overreach near heteroatoms), leading to broad regressions: union-of-all over-tautomerized fused/hetero systems, while smallest-disjoint selection under-selected in cases where chemically correct reachability differs across tautomer forms.
\begin{tcolorbox}[
    colback=green!5,
    colframe=green!40!black,
    left=2pt,right=2pt,top=2pt,bottom=2pt,
    boxrule=0.4pt,
    width=0.92\linewidth,
    enhanced jigsaw,
    breakable,
]
\scriptsize
\begin{Verbatim}[breaklines=true, breakanywhere=true]
// Agent's incorrect approach (conceptual excerpt)
std::vector<Bond*> sortedBonds = mol->bonds();
std::sort(sortedBonds.begin(), sortedBonds.end(), byMinMaxAtomIdx);

for (auto bptr : sortedBonds) {
  if (!startBonds[bptr->getIdx()]) continue;
  boost::dynamic_bitset<> bondsConsidered(mol->getNumBonds()); // reset each seed
  ... build conjSystem ...
  systems.push_back(conjSystem);
}

std::sort(systems.begin(), systems.end(),
          [](auto &a, auto &b){ return a.count() < b.count(); });

for (auto &sys : systems) {
  if (!sys.intersects(bondsToModify)) bondsToModify |= sys;
}
\end{Verbatim}
\end{tcolorbox}

This ``determinism by sorting'' is orthogonal to chemical correctness: the algorithm became stable with respect to indices, but unstable with respect to chemistry, which is what the hash is meant to preserve.

\medskip
{\normalsize\textbf{Significance}}\\
This trajectory illustrates a recurring failure mode in scientific software repair: the model can produce plausible, compilable C++ and even add a targeted reproduction test, yet still fail because the \textit{invariant} is domain-defined rather than purely algorithmic. Here, canonical tautomer hashing depends on nuanced chemical heuristics (functional group exclusions, conjugation density, heteroatom-adjacent overreach prevention). Generic graph/order fixes cannot substitute for those rules, and can silently introduce scientifically meaningful regressions (e.g., incorrectly merging or separating tautomeric systems). The case underscores why ``passes a reproduction'' is insufficient in chemistry-heavy code: correctness is encoded in specialized heuristics that constrain which transformations are chemically admissible.

\end{casebox}

\bibliography{reference}
\bibliographystyle{unsrt}

% \bibliographystyle{plainnat}   % 或 colm 推荐的 style
% \bibliography{refs}

\end{document}